# The Origin of Chondrules: Constraints from Matrix-Chondrule Complementarity

Herbert Palme[1*], Dominik C. Hezel[2,3], Denton S. Ebel[4]

[1]Forschungsinstitut und Naturmuseum Senckenberg, Senckenberganlage 25, D-60325 Frankfurt am Main, Germany

[2]Institut für Geologie und Mineralogie, Universität zu Köln, Zülpicherstrasse 55, Germany,

[3]Natural History Museum, Department of Mineralogy, Cromwell Road, SW7 5BD, London

[4]Department of Earth and Planetary Sciences, American Museum of Natural History, 200 Central Park West, New York, NY 10024-5192, USA

*Corresponding author. E-mail: Palmeherbert@gmail.com

**Abstract**

One of the major unresolved problems in cosmochemistry is the origin of chondrules, once molten, spherical silicate droplets with diameters of 0.2 to 2 mm. Chondrules are an essential component of primitive meteorites and perhaps of all early solar system materials including the terrestrial planets. Numerous hypotheses have been proposed for their origin. Many carbonaceous chondrites are composed of about equal amounts of chondrules and fine-grained matrix. Recent data confirm that matrix in carbonaceous chondrites has high Si/Mg and Fe/Mg ratios when compared to bulk carbonaceous chondrites with solar abundance ratios. Chondrules have the opposite signature, low Si/Mg and Fe/Mg ratios. In some carbonaceous chondrites chondrules have low Al/Ti ratios, matrix has the opposite signature and the bulk is chondritic. It is shown in detail that these complementary relationships cannot have evolved on the parent asteroid(s) of carbonaceous chondrites. They reflect preaccretionary processes. Both chondrules and matrix must have formed from a single, solar-like reservoir. Consequences of complementarity for chondrule formation models are discussed. An independent origin and/or random mixing of chondrules and matrix can be excluded. Hence, complementarity is a strong constraint for all astrophysical–cosmochemical models of chondrule formation. Although chondrules and matrix formed from a single reservoir, the chondrule-matrix system was open to the addition of oxygen and other gaseous components.

Palme, Hezel, Ebel (2015, *EPSL*)

## 1. Introduction.

Chondritic meteorites sample primitive solar system material. Bulk chondrites have not been modified by melting and crystallization processes. Their major element compositions come close to the composition of the Sun (e.g., Lodders et al., 2009). Chondrules are a major structural component of chondritic meteorites. They have experienced temperatures of up to 2000 K with subsequent rapid cooling (Hewins, 1997). The process responsible for making chondrules remains unknown. Suggestions for their formation encompass a wide range of possible mechanisms, such as condensation from a hot solar gas, formation near the Sun and transport with protostellar jets to the asteroid belt and beyond, collisions of molten planetesimals, heating by shock waves, generated by gravitational instabilities in the solar nebula or by supersonic planetesimals, or heating by electromagnetic processes (Ciesla, 2005). A wealth of chemical and isotopic data on single chondrules in a large variety of chondritic meteorites has been collected during the last 40 years (e.g., Jones et al., 2005). Progress has been made in identifying timing and conditions of formation of chondrules. The chondrule formation process is arguably the major unresolved question in cosmochemistry.

Carbonaceous chondrites, a sub-group of chondritic meteorites, contain in addition to chondrules large fractions of matrix, from almost 100% in CI-chondrites to 40% or less in the Renazzo type CR-chondrites (e.g., Fig.1). Many carbonaceous chondrites (e.g., CO, CV) have about equal amounts of chondrules and fine-grained matrix. Matrix is generally crystalline with grain sizes typically below 5μm (Weisberg et al., 2006). In some of the very primitive CR-chondrites matrix is largely amorphous, i.e., non-crystalline (Abreu and Brearley, 2010).

The bulk chemistry of carbonaceous chondrites is defined by the two major components, chondrules and matrix. Although Ca-, Al-rich inclusions (CAI) are an important component in carbonaceous chondrites, their abundance is often overestimated (Hezel et al., 2008). We estimate a maximum contribution of CAIs to bulk meteorite Si of about 2% to 3% and even less for Mg.

Most chondrules in unmetamorphosed carbonaceous chondrites are dominated by FeO-poor olivine (type 1 chondrules). They are particularly prominent in CR-chondrites. The low Si/Mg ratios of chondrules in these meteorites are compensated by high Si/Mg ratios in matrix, leading to chondritic or solar Si/Mg ratios of the bulk meteorites. This complementarity has been demonstrated for the CR-chondrite Renazzo by Klerner (2001), Klerner and Palme (1999) and for other types of carbonaceous chondrites (Hezel and Palme, 2008, 2010; Ebel et al., 2009). Earlier, Wood (1963, 1985) had elaborated on similar relationships based on differences of Fe/Mg in chondrules and matrix of CM-chondrites.

In this paper we focus on the composition of the matrix in carbonaceous chondrites. Since the submission of two earlier publications on this topic by Hezel and Palme (2008, 2010) several papers with precise and careful measurements of matrices in carbonaceous chondrites have appeared in the literature (Wasson and Rubin, 2009, 2010a; Abreu and Brearley, 2010). These studies showed that matrix is compositionally rather uniform and has a distinct chemical signature, characteristic of all carbonaceous chondrites analyzed. In addition, Stracke et al. (2012) have shown that the relatively coarse-grained Allende meteorite is compositionally uniform with regard to Si, Mg and Fe on a scale of a few millimeters. This is significant, because Si, Mg, and Fe plus oxygen make up nearly 90% of the Allende meteorite. Other carbonaceous chondrites should at least be similarly homogeneous. This is also implicitly clear from the generally good agreement of chemical analyses of carbonaceous chondrites by various authors, as discussed below. Here we combine so far scattered data sets to demonstrate the chemical complementarity between chondrules and matrix and discuss possible reasons for it.

Palme, Hezel, Ebel (2015, *EPSL*)

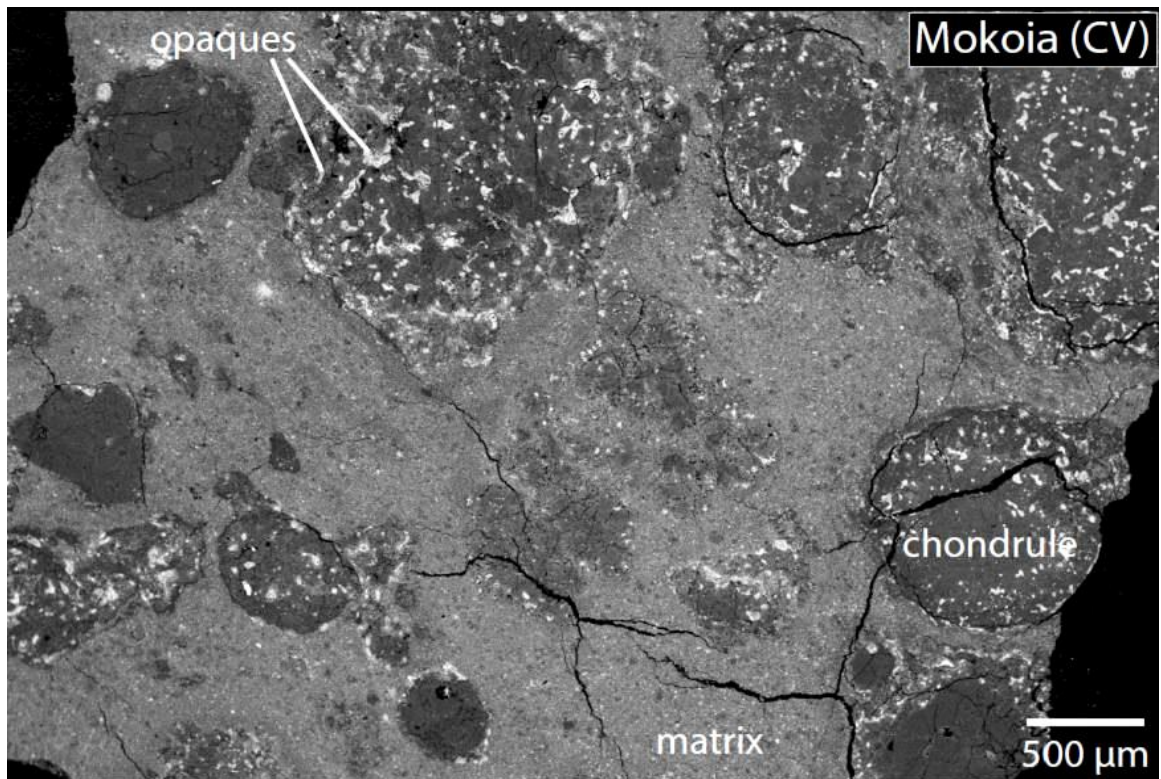

**Figure 1.** Chondrules and matrix in the Mokoia CV-chondrite with backscattered electrons. The darker chondrules indicate lower FeO contents. Bright spots are sulfides or metal.

## 2. Materials and methods

In this paper we use data on bulk compositions, chondrules and matrix of carbonaceous chondrites. In addition to literature data, we report new matrix data of the CV-meteorite Mokoia to complement the data of nearly 100 individual chondrules in Mokoia by Jones and Schilk (2009). Matrix areas from a thin section of Mokoia from the Natural History Museum, London were analyzed with a Cameca SX100 electron microprobe. The accelerating voltage was set to 20 kV, the beam current to 20 nA. The electron beam was defocused to a diameter of 20μm. The following standards have been used (detection limits in wt.%): Si, synthetic fayalite: 0.02; Ti, synthetic rutile: 0.04; Al, synthetic corundum: 0.02; Cr, synthetic chromium oxide: 0.04; Fe synthetic fayalite: 0.04; Mn synthetic $MnTiO_3$: 0.06; Ni synthetic NiO: 0.06; Mg synthetic forsterite: 0.02; Ca, natural wollastonite: 0.03; Na, jadeite: 0.04. The built in PAP-algorithm (e.g., Pouchou and Pichoir, 1991) was used for correction. Typical totals are around 90 wt.%. This is either due to the occurrence of some hydrous minerals and/or a result of the porous nature of the Mokoia matrix. In the latter case absolute element concentrations might be underestimated by up to 10%. Element ratios are unaffected. Iron in metal and sulfide is calculated as FeO, which leads to artificially high FeO concentrations. Hezel et al. (2013) report a metal +sulfide abundance of 3.3 vol.%, and Ebel et al. (2009) only 0.8 area% for Mokoia. A significant contribution to these values is large metal/sulfide grains that were avoided during our matrix measurements. As matrix in Mokoia typically has FeO concentrations of around 35 wt.%, and most opaques phases are sulfide (Hezel et al., 2013), the overestimation of matrix FeO from opaque phases is negligible. The Mokoia matrix data are given in the Supplementary Online Material.

We also discuss the chondrule-matrix relationship in Renazzo. Here we are using recent matrix data from Hezel and Palme (2010) and chondrule data from Klerner (2001). The latter data and analytical details are given in the Supplementary Online Material.

Palme, Hezel, Ebel (2015, *EPSL*)

### 3. Bulk compositions of carbonaceous chondrites are well defined and approximately solar

Wiik (1956) pointed out the compositional uniformity of bulk carbonaceous chondrites. The average carbonaceous chondrite Si/Mg weight ratio of 22 samples analyzed by Wiik (1956) using wet chemical procedures is 1.102 ± 0.034(see compilation by Ahrens, 1965). Later independent wet chemical analyses by Jarosewich (1990) gave 1.109 ±0.036(19 samples). An average carbonaceous chondrite Si/Mg ratio of 1.102 ±0.025(22 samples) was determined by Wolf and Palme (2001) using X-ray fluorescence (XRF). In these data sets there is a small, barely resolvable decrease in the Si/Mg ratio of 3 to 4% in the sequence CI, CM, CV chondrites. Allende (CV) bulk samples have, for example, an av-erage Si/Mg ratio of 1.080 ±0.061(Stracke et al., 2012) slightly below the CI-ratio of 1.12 ±0.05(Lodders et al., 2009), in accord with a Si/Mg ratio of 1.080 ±0.011for an average of a 4 kg Allende sample (Jarosewich et al., 1987). The solar photospheric Si/Mg ratio of 1.10 ±0.2fits with the carbonaceous chondrite ratios, but has larger uncertainties (Lodders et al., 2009).

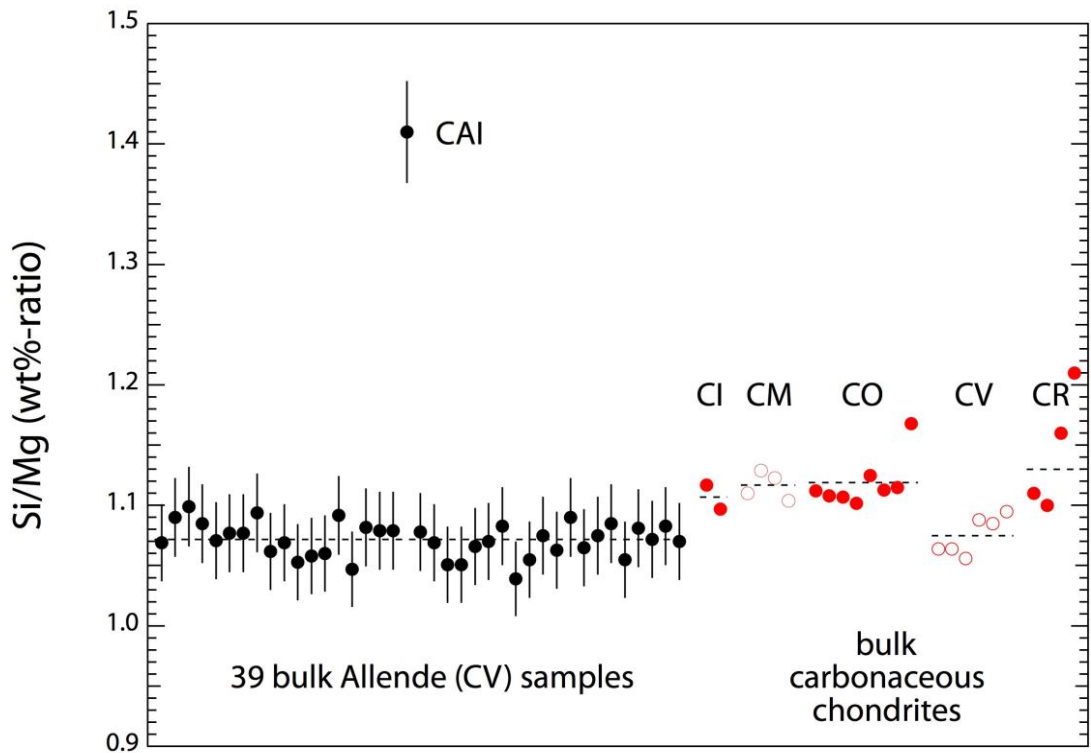

**Figure 2.** Si/Mg ratios of 39 bulk Allende samples, each with an average mass of about 600 mg (Stracke et al., 2012). Only a single sample containing a large CAI deviates from the average. Bulk carbonaceous chondrites of other types are plotted for comparison (Wolf and Palme, 2001).

In Fig. 2 we plot the Si/Mg weight ratios of 39 bulk Allende samples, each with a mass of about 600 mg (Stracke et al., 2012). There is a spread of about 5% within the bulk samples, reflecting analytical uncertainties. Only a single sample containing a large Ca-, Al-rich inclusion has a different Si/Mg ratio. Also shown are bulk carbonaceous chondrite Si/Mg ratios (Wolf and Palme, 2001). For CR-chondrites the data from Wiik (1956) for Renazzo and Al Rais were taken, in addition to two XRF analyses for Renazzo from Klerner (2001). The

Palme, Hezel, Ebel (2015, *EPSL)*

Si/Mg ratios in CR-chondrites range from 1.10 to 1.21, in accord with other carbonaceous chondrites.

Fe/Mg ratios among carbonaceous chondrites are somewhat less constant than Si/Mg ratios. The mean Fe/Mg ratios (by mass) decrease from CI (1.92) through CM (1.77) and CO (1.74) to CV (1.55), according to Wolf and Palme (2001). Kallemeyn and Wasson (1981) report a similar decrease: 1.88 (CI), 1.79 (CM), 1.71 (CO) and 1.64 (CV). The Fe/Mg ratios in the 600 mg Allende samples are, however, as constant as the Si/Mg ratios (Stracke et al., 2012). Thus ratios among the three elements Si, Mg and Fe are constant within a few percent in a single carbonaceous chondrite at a 600 mg sample scale, as demonstrated for Allende, a comparatively coarse grained rock.

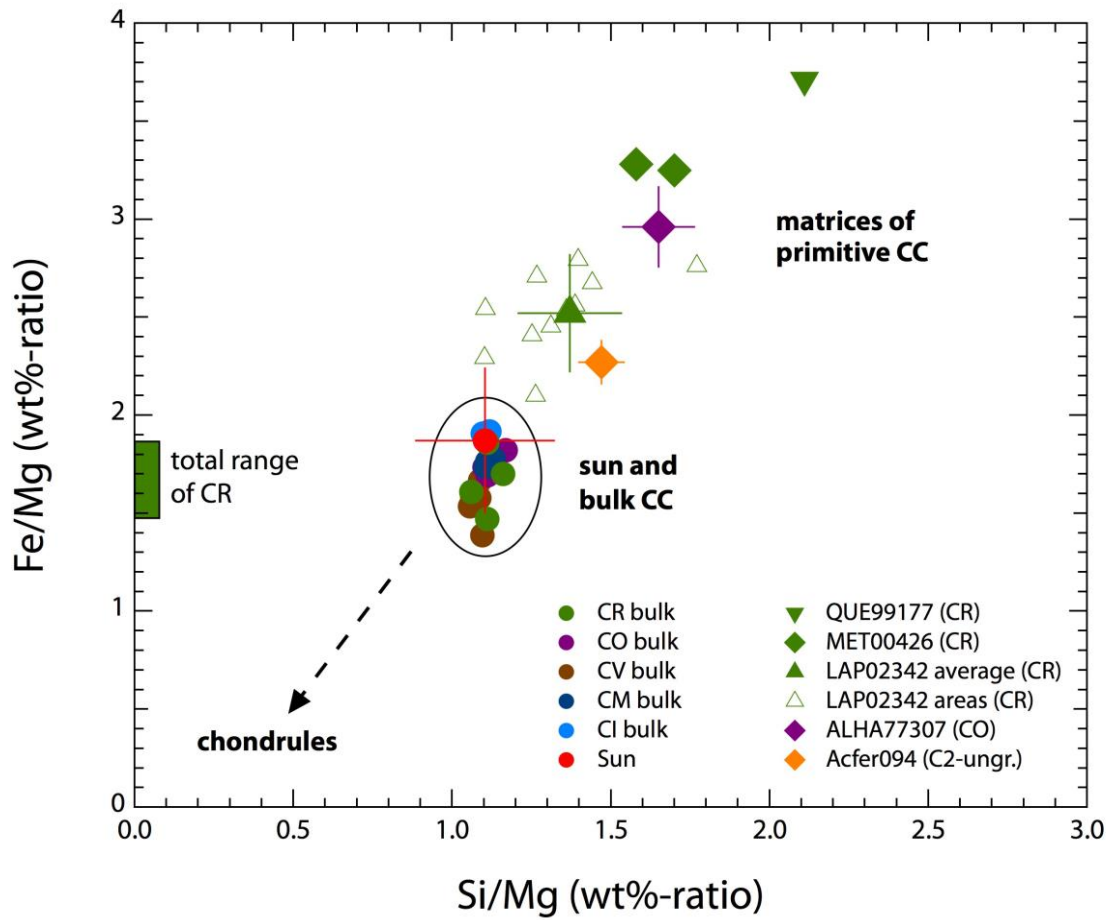

**Figure 3.** Si/Mg vs Fe/Mg ratios in bulk carbonaceous chondrites and in matrices of carbonaceous chondrites. All bulk meteorites except one CV sample plot within error bars of the solar ratios. All matrix samples plotted here have considerably higher ratios. These primitive carbonaceous chondrites have roughly equal amounts of chondrules and matrix. Some CR bulk analyses by instrumental neutron activation analysis (INAA) are without Si, so their range in Fe/Mg ratios is indicated at the ordinate (Kallemeyn and Wasson, 1994; Bischoff et al., 1993).

**4. Composition of matrix in carbonaceous chondrites**

We compare in Fig.3 the Fe/Mg and Si/Mg ratios of bulk carbonaceous chondrites with ratios measured in the matrices of carbonaceous chondrites. The bulk carbonaceous chondrite data were taken from Wolf and Palme (2001). For CR chondrites there are more data for Fe/Mg than for Si/Mg ratios, because Si cannot be determined with instrumental neutron activation analysis. We have indicated in Fig.3 the range of Fe/Mg ratios as determined by Kallemeyn et al. (1994) and Bischoff et al. (1993). In matrix analyses we have focused on recent analyses of very primitive carbonaceous chondrites. Acfer 094, for example, has very high contents of presolar SiC and nano-diamond grains, shows extreme disequilibrium among silicates and contains a high fraction of amorphous material (Wasson and Rubin, 2010a). The matrix analysis of Acfer 094 plotted in Fig.3is based on a large number of rather similar analyses with a mean Si/Mg ratio of 1.50 ±0.08and a mean Fe/Mg ratio of 2.36 ±0.012(Wasson and Rubin, 2010a). A careful study of matrix in the CR chondrite LAP 02342 showed some variability in matrix compositions (Wasson and Rubin, 2009). In Fig.3we plot the results of individual analyses of 11 matrix areas about 50 ×50μmeach, as well as the total average. All analyses plot significantly above the ratios for bulk Si/Mg and Fe/Mg ratios of carbonaceous chondrites. The two recently identified very primitive CR chondrites MET 00426 and QUE 99177 have largely amorphous matrix considered by Abreu and Brearley (2010) to be a primary feature and not the result of aqueous alteration. Averages of analyses of five regions of interchondrule matrix and eight regions of fine-grained chondrule rims of MET 00426 and three regions of matrix of QUE 99177 are plotted in Fig.3. All areas have high Si/Mg and Fe/Mg ratios. As emphasized by Abreu and Brearley (2010) there is no significant difference between inter-chondrule matrix and matrix rimming chondrules. The matrix of ALHA 77307, a primitive CO chondrite is similarly high in Si/Mg and Fe/Mg (Brearley, 1993).

The matrices of other type 3 and type 2 carbonaceous chondrites have similar signatures, as documented by Hezel and Palme (2010) and other literature data. For example, average matrix compositions using the data of Hezel and Palme (2010) for the four carbonaceous chondrites Efremovka (CV), Renazzo (CR), Kainsaz (CO) and El-Quss Abu Said (CM) give Si/Mg ratios from 1.33–2.05 and Fe/Mg ratios from 1.68–3.09. The compilation of matrix analyses by Huss et al. (2005) also shows high Si/Mg and Fe/Mg ratios. Out of 23 matrix analyses of carbonaceous chondrites, 20 have Si/Mg ratios above the bulk meteorite ratios. Data reported by Clarke et al. (1971) on various fractions of Allende show the same picture. A large bulk Allende sample has a Si/Mg ratio of 1.08. For a 2.6 g matrix sample analyzed by the same authors a ratio of 1.20 was determined. Thus there is no doubt that the composition of matrices of carbonaceous chondrites is different from the bulk compositions of their parent meteorites. It appears that the difference is largest for the most primitive CR meteorites, as displayed in Fig.3. For other groups of carbonaceous chondrites the difference is less pronounced but nevertheless real. Data for matrix in ordinary chondrites also fall on the trend given in Fig.3, as demonstrated by Huss et al. (1981, 2005) and additional data by Grossman and Brearley (2005).

In Fig.4 the data of Hezel and Palme (2010) on matrix and chondrules of several CV-chondrites are plotted in the Fe/Mg vs. Si/Mg diagram of Fig.3. Individual matrix analyses show a wide compositional spread, although the data fall within the range of matrix analyses of the pristine CR-chondrites of Fig.3. The larger spread for individual analyses in Fig.4 has two causes: (a) The analyses in Fig.4 from Hezel and Palme (2010) were done with a10μm electron beam. Each point in Fig.4 represents a single analysis. Abreu and Brearley (2010) used averages of 6 to 35 analyses for each individual matrix region with an EMP beam size of 10μm (Fig.3). Wasson and Rubin (2009, 2010a) used small beam sizes, but report averages of areas of around 50μm×50μm. (b)The meteorites plotted in Fig.4 are less primitive than those plotted in Fig.3. Although parent body alterations produce no changes on a cm scale (see Stracke et al., 2012), they may cause significant redistributions on a μm scale. The presence of Ca-, Fe-silicates indicates major redistribution of Ca (e.g., MacPherson and Krot, 2014 and references therein). The average Ca content calculated from 150 Mokoia matrix analyses is 1.47 ±3.9 compared to 10.92 ±2.2 for Mg (Supplementary Online Material). The mean Fe/Mg ratio of the Mokoia matrix analyses is 2.58 ±0.58, clearly above the bulk meteorite value of

Palme, Hezel, Ebel (2015, *EPSL*)

1.67 (Kallemeyn and Wasson, 1981). The average Si/Mg ratio is 1.47 ±1.84, as high as expected but much less well defined. However, out of 150 matrix analyses only 19 have Si/Mg ratios below the average carbonaceous chondritic ratio of 1.10 (see Fig.5).

The matrix of QUE 99177 (CR2) is significantly higher in Fe/Mg and Si/Mg than the MET 00426 (CR2) matrix (Fig. 3 in this paper and Fig. 7 in Abreu and Brearley, 2010). Both meteorites are considered the most primitive members of the CR-meteorite group. Also, the difference in matrix composition between Acfer 094 (Wasson and Rubin, 2010a) and LAP 02342 (Wasson and Rubin, 2009) is real (see also Fig. 3). The matrix analyses of CM, CO, CR and CV (red.) meteorites in Fig.4 seem to occupy a separate field for each meteorite type.

Fig.5 demonstrates the chemical complementarity of chondrules and matrix for the Mokoia (CV oxidized) carbonaceous chondrite. The large spread in chondrule compositions is typical of most carbonaceous chondrites. The spread in matrix analyses is discussed above.

On a macroscopic scale matrix is compositionally rather uniform in a single carbonaceous chondrite. Matrix-filled areas be-tween chondrules and matrix rimming chondrules have essentially the same chemical compositions. Larger variations occur on a sub-millimeter scale (Wasson and Rubin, 2009; Brearley, 1993). In other cases, parent body weathering has obscured the original homogeneity, in particular with regard to CaO and $Na_2O$, which are strongly susceptible to alteration, as discussed above for CaO (Zolensky et al., 1993).

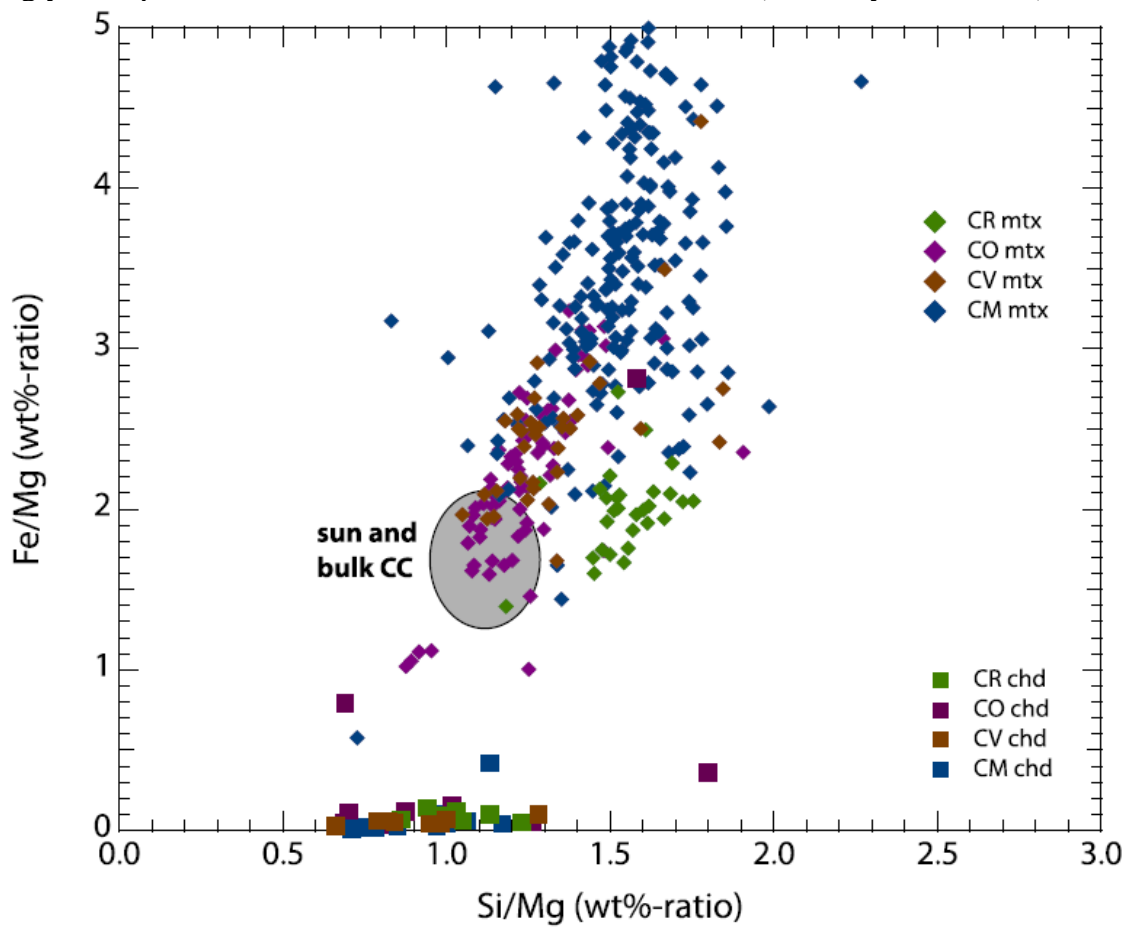

**Figure 4.** Same diagram as Fig.3 with data on matrices and chondrules of CR, CO, CV and CM chondrites from Hezel and Palme (2010). Most matrix analyses have higher Fe/Mg and Si/Mg ratios than bulk carbonaceous chondrites. The corresponding chondrule analyses are lower in both ratios. The comparatively low Fe/Mg ratios of matrix of CR-chondrites (Renazzo) is the result of avoiding metals when analyzing matrix.

Palme, Hezel, Ebel (2015, *EPSL*)

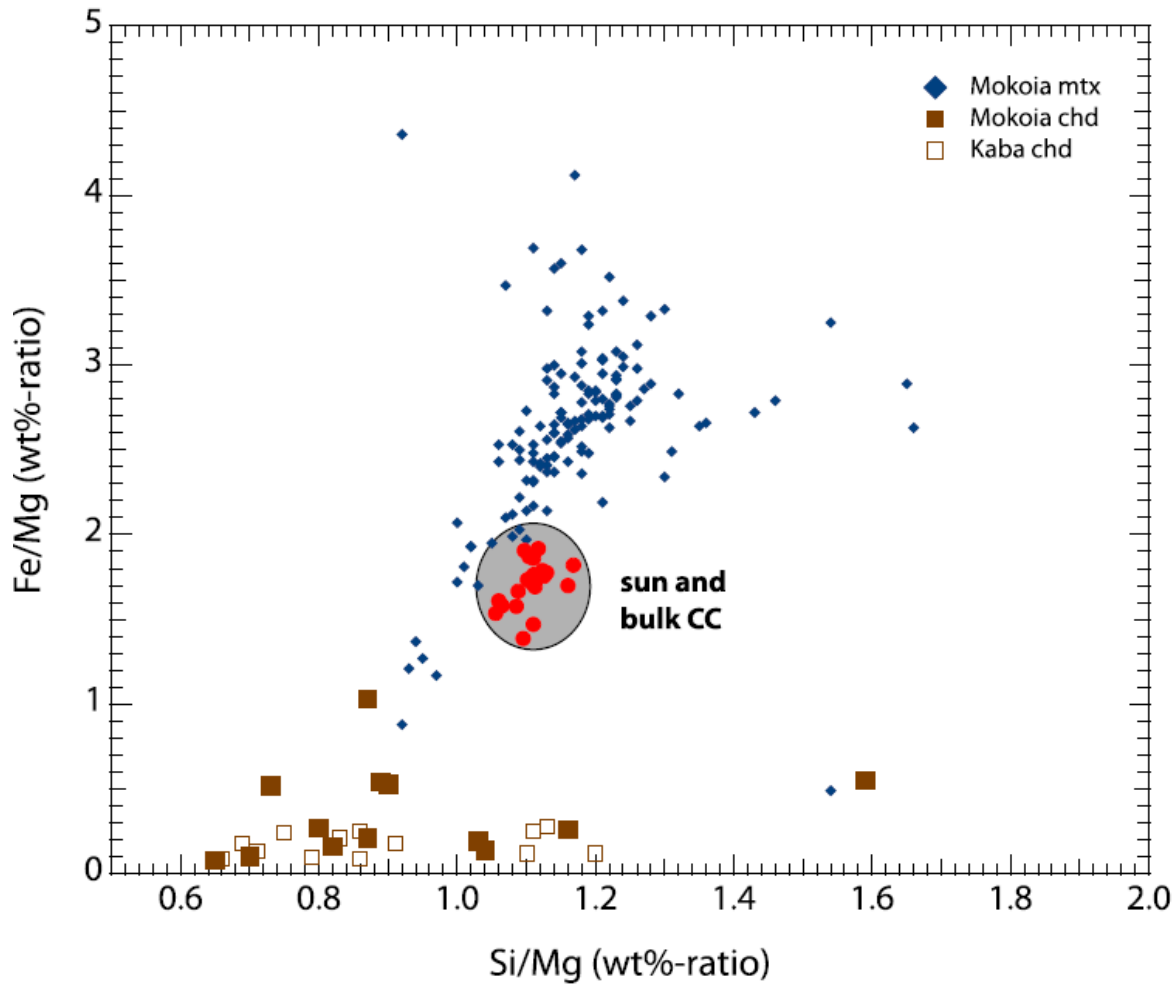

**Figure 5.** The complementary Si/Mg and Fe/Mg relationships of chondrules and matrix in the Mokoia (CV) carbonaceous chondrite. Matrix data are in the Supplementary Online Material. Chondrule data are from Kimura and Ikeda (1998). Chondrules from Kaba (CV) are included for comparison. Data for Mokoia chondrules by Jones and Schilk (2009) cannot be plotted, because there are no data on Si.

## 5. Composition of chondrules in carbonaceous chondrites

Since matrix and chondrules are the two volumetrically relevant components in carbonaceous chondrites, the chondrule component must compensate the high Fe/Mg and Si/Mg ratios of matrix, as indicated in Figs.3–5.

Bulk chemical data for chondrules do indeed indicate low Si/Mg ratios of whole chondrules. Most of the chondrule compositions reported by Hezel and Palme (2010) have lower than chondritic Si/Mg ratios, as expected (Fig.4). Chondrule compositions for Renazzo determined by Klerner (2001) also have generally low Si/Mg and Fe/Mg ratios and are thus complementary to matrix compositions (Supplementary Online Material). Out of 92 chondrule analyses from Mokoia by Jones and Schilk (2009), 91 have a sub-chondritic Fe/Mg ratio. Eleven out of thirteen bulk chondrule analyses in Mokoia (CV) by Kimura and Ikeda (1998) show lower than bulk meteorite Si/Mg and Fe/Mg ratios. Chondrules in the Renazzo meteorite have been shown by 3-dimensional analysis to be complementary to each other in Fe/Si and Fe/Mg ratios (Ebel et al., 2008), i.e., average Fe/Mg and Si/Mg ratios of chondrules are al-ways below the chondritic or bulk meteorite ratios.

Palme, Hezel, Ebel (2015, *EPSL)*

Another example of the complementary relationship between chondrules and matrix is shown in Fig.6 where we plot Fe against Mg in chondrules and matrix of the Mokoia CV-chondrite. The data for 92 chondrules are from Jones and Schilk (2009) and those for 150 matrix analyses are from the Supplementary Online Material. The variation in the bulk chondrite Fe/Mg ratios is comparatively small compared to the enormous variations in Fe/Mg ratios of individual chondrules. Since chondrule data were obtained by instrumental neutron activation analysis (INAA) of whole chondrules, they include none of the uncertainties inherent in 2D analyses of random sections through heterogeneous chondrules (Hezel and Kießwetter, 2010). Only very few of the chondrules fall in the field of matrix compositions and very few matrix analyses plot with the chondrules. Chondrule compositions range from nearly pure forsterite $Mg_2SiO_4$ (34.55 wt.% MgO) to increasingly FeO-rich, with some reaching bulk meteorite compositions. The array suggests initial formation of Mg-rich chondrules at high temperatures from a chondritic reservoir, with increasing FeO in olivine at lower temperatures. At some point chondrule formation stops and the residual FeO-rich material ends up as fine-grained matrix. Plots similar to Fig.4 for Mokoia can also be generated for Allende.

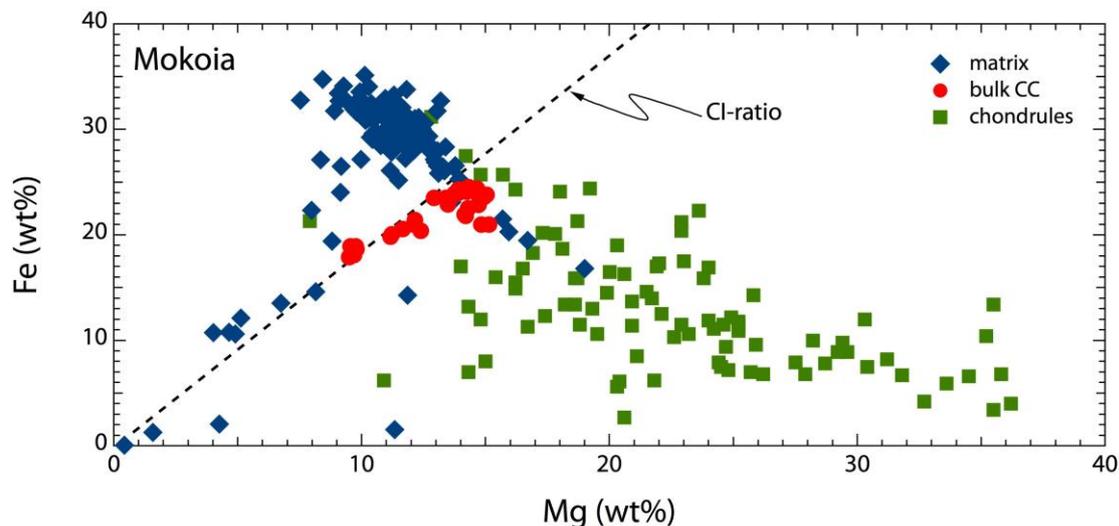

**Figure 6.** A plot of Fe vs Mg of bulk carbonaceous chondrites and of chondrules and matrix of the Mokoia (CV) meteorite. Matrix is complementary to chondrules. Matrix and chondrules of Mokoia are affected by hydrous and anhydrous alteration (Kimura and Ikeda, 1998). However, bulk chondrules show no signature of alteration (Jones and Schilk, 2009). Bulk carbonaceous chondrite data are from Wolf and Palme (2001), chondrule data from Jones and Schilk (2009). For matrix data see supporting online material.

**6. Fractionation of refractory elements**

The two CV chondrites Allende and Y-86751 have almost identical bulk chemical compositions including chondritic (solar) Ca/Al ratios. But the distribution of Ca and Al between matrix and chondrules is very different in the two meteorites (Hezel and Palme, 2008). Contrary to Allende, the Y-86751 CV-chondrite has significantly higher than chondritic Ca/Al ratios in chondrules and a low ratio in matrix. The reason for the unusually high Al in matrix of Y-86751 is a large population of tiny spinel grains (Murakami and Ikeda, 1994). The excess Al in the matrix is compensated by un-usually low Al in chondrules. Both components combined produce a chondritic, i.e., solar bulk Ca/Al-ratio.

Another strong matrix-chondrule relationship involving refractory elements is apparent from the super-chondritic Ti/Al ratios in chondrules of Renazzo (Klerner and Palme, 1999; Klerner, 2001). Titanium is preferentially incorporated into chondrules in these meteorites, i.e., complementary to a Ti deficit in the matrix. In Fig.7, we show the Al–Ti relationship for chondrules and matrix in the Renazzo meteorite. Despite significant variations in Ti and Al contents of individual chondrules there is a clear trend toward super-chondritic Ti/Al ratios in chondrules. The elevated Ti/Al ratios in chondrules are important because Ti and Al are both immobile refractory elements. Their redistribution on the CR parent body would require thermal and/or aqueous processes sufficiently intense to destroy the presolar grains found in Mokoia and Renazzo matrix (Huss et al., 2003).

A similar figure as for Renazzo is shown for Mokoia in Fig.8. A clear super-chondritic Ti/Al ratio is present in the Jones and Schilk (2009) data. The error bars are 15% for Ti, which is the mean of the range of uncertainties ascribed to the Ti analyses by Schilk (1991). The corresponding uncertainty for Al is 3%. There are no error bars for matrix analyses, as the spread of Al and Ti concentrations is primarily the effect of inhomogeneities in the matrix on a scale of 10μm. Although chondrules have on average a super-chondritic Ti/Al ratio the compositional spread in matrix analyses is too large to uniquely define a subchondrtic Ti/Al ratio for matrix.

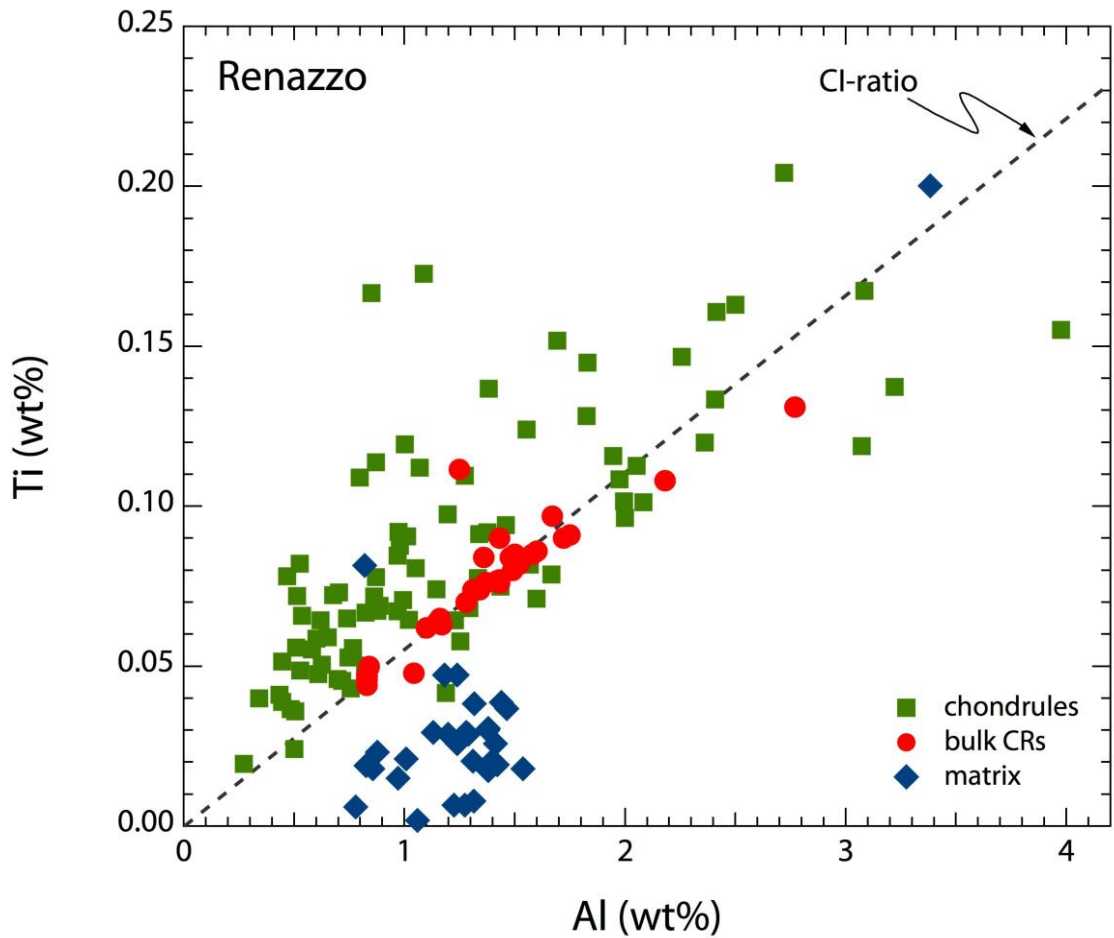

**Figure 7.** Ti vs Al in bulk CR carbonaceous chondrites, Renazzo chondrules and matrix. A complementary relationship is evident. The high Ti content of chondrules may reflect preferred incorporation of condensed perovskite. Data are from (Klerner, 2001; Klerner and Palme, 1999; Wolf and Palme, 2001).

Palme, Hezel, Ebel (2015, *EPSL)*

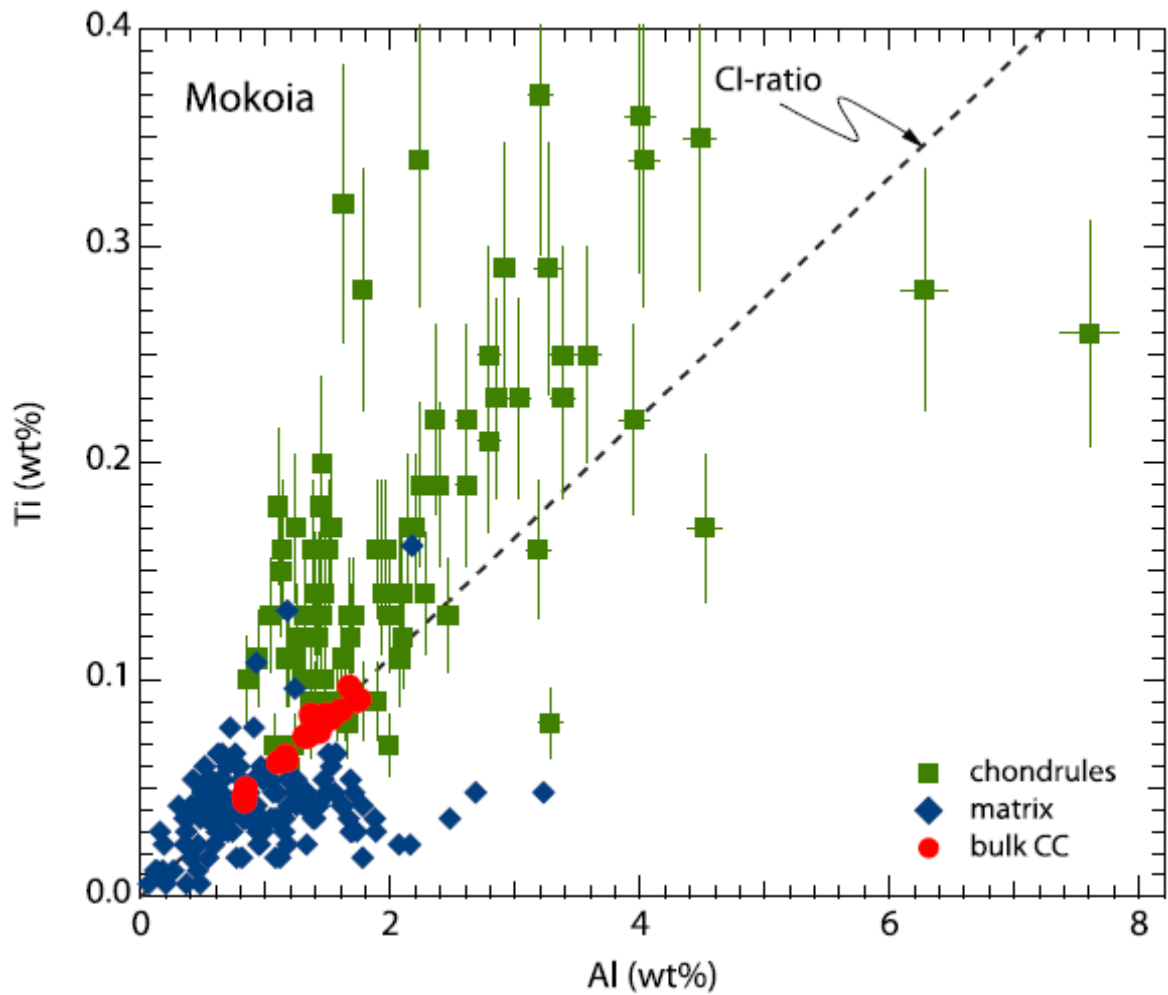

**Figure 8.** Ti vs. Al in chondrules and matrix of Mokoia. The data for chondrules are from Jones and Schilk (2009). The error bars are set at 15%, the mean of the range given by Schilk (1991). Chondrules have a super-chondritic Ti/Al ratio. The complementary nature of matrix is not clear, because of the large variability, probably caused by weathering of matrix.

**7. Compositional differences between chondrules and matrix were established before accretion**

*7.1. Si/Mg ratios*

The complementary relationship between chondrules and matrix would be trivial had it resulted from elemental redistribution during thermal metamorphism or aqueous alteration on the carbonaceous chondrite parent bodies. It is conceivable that the high Si/Mg ratio of matrix in carbonaceous chondrites is the result of aqueous alteration of chondrule mesostases. Leaching of chondrule glass may lead to loss of Si from chondrules. Silicon would then be transported with aqueous fluids into matrix producing the observed low Si/Mg of chondrules and the high Si/Mg of matrix. Grossman et al. (2002) noticed that mesostasis in Semarkona, one of the most primitive ordinary chondrites, is often zoned, with depletion of Ca at the rim. In some cases Si shows a similar depletion to Ca, in other cases Si shows the opposite effect or no zoning. It is unclear if and to what extent Si is affected by the hydration of chondrules.

Ikeda and Kimura (1995) and Kimura and Ikeda (1995) studied the groundmass (i.e., mesostasis) in chondrules of the Allende meteorite and identified high temperature processes that lead to loss of Si from mesostasis. Unaltered mesostasis is either glassy or microcrystalline with plagioclase or normative plagioclase as the major component. Kimura and Ikeda (1995) describe the decomposition of the albite component to nepheline with the release of $SiO_2$, which could lead to loss of Si from chondrules. Ikeda and Kimura (1995) and Kimura and Ikeda (1995) believe that these alteration processes occurred in the solar nebula, before parent body accretion. Krot et al. (1995, 1998) suggest that the observed alteration features are the result of parent body processes at relatively low temperatures (<300°C) by fluid–rock interaction. In this case $SiO_2$ lost from chondrules could enter matrix.

We assume here that the proposed redistribution of $SiO_2$ occurred on the parent body and estimate the effect such a process could potentially have on the chondrule and matrix Si/Mg ratios.

The decomposition of albite in chondrule mesostasis and its effect on the Si/Mg ratio was discussed by Hezel and Palme (2010), who concluded that this process is not capable of producing the observed Si/Mg fractionation. Here we present a more detailed estimate. The content of $SiO_2$ in mesostasis of chondrules is between 50 and 60%, according to data by Berlin (2009) on primitive CR and CO chondrites and on data of the mesostasis in Semarkona (LL3.0) by Grossman et al. (2002). For our calculations we will use 60 wt.%. It is also necessary to know the fraction of mesostasis in chondrules. Assuming that (a) all Al of a chondrule is in mesostasis (b) mesostasis has 20% $Al_2O_3$(albite) and (c) bulk chondrules have 3 wt.% $Al_2O_3$ we calculate a maximum fraction of about 15% mesostasis in chondrules.

Ikeda and Kimura (1995) determined the composition of unaltered and altered mesostasis of Allende chondrules. The difference between them is about 6 wt.% $SiO_2$ lost during anhydrous alteration. We will use 10%. With 50 wt.% $SiO_2$ in altered mesostasis, 15% mesostasis in chondrules and 50 wt.% bulk chondrule $SiO_2$, altered mesostasis will contribute 7.5 wt.% SiO2to the bulk chondrule inventory of $SiO_2$, compared to 9% for unaltered mesostasis. Thus chondrules lose a maximum of 1.5 wt.% $SiO_2$ or 0.7 wt.% Si if originally glassy or microcrystalline mesostasis is totally altered by anhydrous processes. This decrease of $SiO_2$ in chondrules would produce a decrease of the bulk chondrule Si/Mg ratio by 3%, assuming Mg is unaffected by alteration (35 wt.% MgO for bulk chondrules). How does this affect the Si/Mg ratio of matrix? Wasson and Rubin (2010b) have studied the matrix in the very primitive meteorite Acfer 094. They report an average of 9.77 wt.% Mg. A chondritic Si/Mg ratio would require 10.75% $SiO_2$ with a chondrule/matrix ratio of 1, matrix would gain 0.7 wt.% Si leading to a Si/Mg ratio of 1.17, which is far lower than the observed Mg/Si ratio of 1.50 (Wasson and Rubin, 2010a). The calculated effect is, however, an upper limit as in the calculation we have assumed extreme values. Furthermore, alteration does not affect the mesostasis of all chondrules in a meteorite (Brearley, 2003). Even Allende, considered to be heavily altered, contains glassy chondrules (Ikeda and Kimura, 1995). These authors classify from a total of 11 chondrule mesostases of Allende, 4 with intense, 4 with medium and 3 with slight alteration. The assumption that alteration has affected all chondrules is thus not realistic and the increase in matrix Si/Mg caused by alteration should be considerably lower than the upper limit estimated here.

The observed enrichment of Si in matrix is largest in the most primitive CR chondrites (Abreu and Brearley, 2010). Berlin (2009) reported that mesostasis in 20 chondrules of MET00426 is glassy, thus not affected by alteration. Yet, matrix in this meteorite has a Si/Mg ratio of 1.7 (Fig.3). Hezel and Palme (2010) found no differences in Si depletion of chondrules and Si enrichment of matrix between reduced and oxidized CV-chondrites. The more extensive alteration of chondrules in oxidized CV-chondrites does not produce a stronger effect.

If aqueous alteration had modified the chemistry of chondrules one would expect a particularly strong effect on some trace elements such as U, Sr and Ba which are extremely sensitive to the influence of water in terrestrial environments. Rocholl and Jochum (1993) have demonstrated low temperature redistribution of U in carbonaceous chondrites. Effects on Ba and Sr are also present, although to a much smaller extent. All three elements, U, Sr and Ba are incompatible and are strongly concentrated in chondrule mesostasis (Alexander, 1994). Exposure of mesostasis to water should readily remove these elements from chondrules and release them to matrix more readily than Si. The average Sc (an immobile element) to U ratio of 15 Allende chondrules analyzed by Gerber (2012) is 740 ±180, essentially identical to the Sc/U CI ratio (chondritic) of 717 (Palme et al., 2014b). The average Sr/Ba ratio of the same set of data is 3.40, compared to the CI ratio of 3.35. Bland et al. (2005) have studied matrices of 22 carbonaceous chondrites with laser ablation techniques. They conclude that less than 1% of Sr was transferred from chondrules to matrix in Allende. Even the heavily altered CM meteorites Nogoya, Murchison and Cold Bokkeveld do not show enrichment of Sr or U in their matrices. Bland et al. (2005) conclude that 'aqueous alteration did not produce chondrule/matrix fractionation of trace elements'.

All CR-chondrites have suffered aqueous alteration based on the presence of phyllosilicates. In addition, all CR-chondrites contain abundant metal. Wasson and Rubin (2010a) stated: 'The aqueous alteration resulted in only minimal oxidation of coarse metal.' It appears highly unlikely that aqueous alteration has leached Si from mesostases of chondrules and that at the same time metal was largely unaffected by water.

In summary, the observed low Si/Mg ratios of chondrules and the high ratios in matrices of carbonaceous chondrites is not produced by alteration of chondrules inside a parent planetesimal. A small reduction of the Si/Mg ratio of chondrules by anhydrous alteration may be possible. But even if the effect is real, it is so small that it cannot provide the Si required for the high Si/Mg ratio in matrix. Since water has not removed U, Ba or Sr from mesostasis leaching of Si seems impossible.

*7.2. Fe/Mg ratios*

The enhanced Fe/Mg ratios of matrix (Figs.3–5) in carbonaceous chondrites could be explained by ejection of liquid metal droplets from rotating chondrules as suggested by Grossman and Wasson (1985). If chondrules had initially CI (chondritic) Fe/Mg ratios, a massive loss of metal was required to achieve the observed low FeO-contents of chondrules. Palme et al.(2014a) reported essentially chondritic Ni/Co ratios in 200 chondrules of CV-chondrites analyzed by several groups. Chondritic Ni/Co and, with a somewhat larger spread, Fe/Ni ratios in chondrules limit the ex-tent of metal loss during chondrule melting. As Ni is much more siderophile than Co, any liquid metal that forms during chondrule melting will have high Ni/Co and Ni/Fe ratios. If this metal is expelled from chondrules, the residual chondrule will be deficient in Ni relative to Co and Fe, destroying the chondritic Ni/Co ratios in chondrules. Significant loss of metal during chondrule melting is thus impossible. Only tiny metal grains (µm sized) could be ejected without destroying the observed chondritic ratios among Fe, Ni and Co in the residual chondrules (Palme et al., 2014a).

The present distribution of FeO between matrix and chondrules in Mokoia (Figs.5, 6) also cannot be the result of parent body alteration. Metamorphic redistribution will tend to equilibrate FeO-poor olivine with FeO-rich olivine with the tendency to produce uniform olivine compositions. The opposite, production of refractory, FeO-poor olivine by Fe–Mg exchange between chondrules and matrix is thermodynamically impossible The FeO enrichment of the rims of forsteritic olivine occurs within a few µm in many carbonaceous chondrites, forbidding cation diffusion of more than a few µm on the CV-parent body (e.g., Hua et al., 1988) and thus strongly limiting cation redistribution.

There is little doubt that the complementary relationships among major and trace elements identified above were established before parent body accretion and thus require the formation of chondrules and accretion of complementary matrix from a single reservoir of solar composition as discussed in detail by Hezel and Palme (2008, 2010). The complementary relationship between refractory elements Al, Ca and Ti in some chondrites demonstrates that complementarity cannot be explained by a volatility related process, i.e., evaporation from chondrule melts at high temperatures will not produce these fractionations.

## 8. Implications for existing models of chondrule formation

The main thrust of chondrule formation models has been to find an appropriate source of energy for melting of dust clumps, thought to be chondrule precursors. The presence of matrix is often not considered in these models. One reason is that the majority of meteorites in our collections, the ordinary chondrites (OC), have only ~15 vol.% matrix (Grossman and Brearley, 2005). Matrix composition has a small effect on the bulk OC meteorite composition, notwithstanding the fact that OC matrix has chemical characteristics very similar to carbonaceous chondrite matrix (Zolensky et al., 1993; Huss et al., 1981). Chondrite formation models that involve independent sources of matrix and chondrules, for example from different formation locations in the solar nebula, cannot account for the observed complementarity.

For example, in stellar wind models (e.g., Shu et al., 2001) chondrules come from the inner edge of the accretion disk near the Sun, where small grains, i.e., chondrules, could be entrained in gas and transported away from the Sun. As grains are carried away from the Sun some may fall back on the protoplanetary disk and mix with cold, local dust to produce chondrites. The relative proportions of matrix and chondrules mixed in such a process is unconstrained. Both non-chondritic components must mix in exactly the right proportions to yield a chondritic bulk composition. Fractionations of refractory elements are particularly difficult to explain in any model requiring mixing of chondrules from <1AU with cold matrix from >3AU in the protoplanetary disk.

Collision of molten planetesimals has recently come back into fashion as a way of making chondrules (Zook, 1981; Lugmair and Shukolyukov, 2001; Asphaug et al., 2011). One model of the impact physics is provided by Asphaug et al. (2011) where ~30–100km diameter, incompletely differentiated, molten planets collide at velocities comparable to their two-body escape velocity ~100m/s, leading ultimately to the formation of millions of glass spheres, that might at appropriate cooling rates be "chondrules". Once formed, these chondrules may encounter fine-grained dust and the assemblage could ultimately form a small planetesimal. The amount of matrix accreted to chondrules is not specified in the model. The metal cores of the initial planetesimals, however, must be converted to micrometer-sized fragments that then must be oxidized and mixed with midplane dust to be available when chondrules accrete iron rich dust to produce a roughly chondritic bulk Fe/Mg ratio (see Fig.3). A model by Sanders and Scott (2012) based on a concept similar to that of Asphaug et al. (2011) encounters the same difficulties in explaining roughly chondritic Fe/Mg ratios of bulk carbonaceous chondrites.

Recently, it has been proposed that highly local high temperatures can result from magneto-rotational instabilities (MRI) in the disk (McNally et al., 2013). Such models are in principle compatible with the chemical constraints imposed by complementarity. MRI-induced current sheets are predicted to be ubiquitous where ionization is present. Current sheets provide a disk-wide mechanism for heating small regions to chondrule melting temperatures, while preserving nearby matrix grains.

Models of chondrule formation by nebular shocks (Desch et al., 2005; Connolly and Love, 1998) are also compatible with matrix chondrule complementarity. It has been pointed out by Desch et al. (2005) that nebular shock will heat fine grained particles much less effectively than mm-sized chondrules or chondrule precursors. These authors conclude that "shock wave heating can explain the accretion of rims on chondrules and the complementary relationship between rim and chondrule volatile chemistries". In such models it is assumed that mm-sized dust balls and µm-sized volatile rich matrix grains are simultaneously present in the pre-shock nebula.

This, however, does not explain the large range of bulk chondrule compositions in carbonaceous as well as in ordinary chondrites (Jones et al., 2005; Hezel et al., 2006; Hezel and Palme, 2007). Several hundred chondrules with individual masses in the milligram range and of variable chemical compositions produce chondrites with chemical compositions constant at the gram scale (Jarosewich, 1990; Stracke et al., 2012).

An important consequence of the matrix-chondrule complementarity is that the major fraction of matrix in carbonaceous chondrites is unlikely to be of interstellar origin. The interstellar medium is thought to represent the solar composition, with non-volatile elements residing quantitatively in grains (Savage and Sembach, 1996). The enrichment of Si and Fe in matrix is the result of removal of Mg-rich chondrule material from a common reservoir, leading to the non-chondritic composition of matrix.

## 9. A closed system for chondrule and matrix formation?

In the preceding section we argued that chondrules and matrix formed from a single reservoir. Our arguments are based on mass balance of major and minor element abundances in matrix and chondrules. However, the reservoir may be open to volatile elements and other minor components with negligible effects on the mass balance of major elements.

The oxygen isotopic composition of chondrules in CV-chondrites is quite variable and very different from matrix. At least three reservoirs are required to produce these variations. Some chondrules even contain relict minerals with very different oxygen iso-topic composition than the main part of the chondrule (Clayton et al., 1983; Jones et al., 2004; Rudraswami et al., 2011). The chondrule formation region was apparently open to oxygen.

One possibility for delivery of small stable isotope anomalies are Ca-, Al-rich inclusions (CAI). Individual chondrules in ordinary and carbonaceous chondrites contain refractory elements at widely different abundance levels (Grossman and Wasson, 1983; Rubin and Wasson, 1987). Each chondrule precursor may have sampled a different CAI-component. There is evidence for the relative movement of refractory grains and gas in the nebular reservoir where chondrules and matrix formed. From the Allende reservoir a super-refractory component must have separated early before formation of the main part of Allende, as bulk Allende has slightly fractionated refractory elements, suggestive of prior removal of a super-refractory component (Stracke et al., 2012 and references therein).

These examples demonstrate that despite a complementary relationship of chondrule and matrix, the system may be open for low temperature components. In addition, the system parental to chondrules and matrix is not necessarily homogeneous. Isotope anomalies introduced by refractory components may be preserved through chondrule formation. Unusual trace element data in individual chondrules indicate that the nebular reservoir from which chondrules and matrix formed was not chemically uniform (e.g., Misawa and Nakamura, 1988).

## 10. Complementarity, a general principle for chondritic meteorites

The concept of complementarity is applicable to chondritic meteorites in general.
Here we will give an example without going into detail. Ordinary chondrites have well defined bulk compositions. H-chondrites, for example, have an average bulk Fe/Mg ratio of 1.95 (Jarosewich, 1990), identical to the CI ratio of 1.95 (Palme et al., 2014a) and within the range of the solar ratio (2.00 ±0.4; Lodders et al., 2009). More than half of the total Fe (58.2%) of H-chondrites is in metal, 12.6% in sulfide and 29.2% in silicates. In addition, lithophile non-volatile elements occur in approximately solar element ratios. The silicate carrier of ordinary chondrites consists predominantly of chondrules with extremely variable compositions (e.g., Jones et al., 2004; Hezel et al., 2006; Hezel and Palme, 2007). Yet a few hundred chondrules plus associated metal and sulfide define, on a gram scale, the solar Fe/Mg ratio. Clearly, chondrules, metal and sulfides must have formed from a single chondritic reservoir (see also Ebel et al., 2008). The independent addition of a metal component is not plausible.

The average Fe/Mg ratio (wt.%) of bulk R-chondrites is 1.89 (Bischoff et al., 2011). Chondrules in type 3 lithologies of the highly oxidized R-chondrites have similarly low FeO contents as chondrules in type 3 ordinary chondrites (Kita et al., 2013). In R-chondrites the complementary Fe-rich phase is matrix (ca. 50 vol.%) which contains the Fe required to produce the solar bulk Fe/Mg ratio (Bischoff et al., 2011). In H-chondrites this phase is FeNi-metal. In H3-chondrites 90% of the silicates is consumed in the formation of, on average, FeO-poor chondrules. The remaining Fe ends up as FeNi-metal. In R-chondrites chondrule formation stopped before all silicates were consumed, the leftover Fe ended up as FeO in matrix. Although the initial stages of chondrule formation may have been very similar in R- and in H-chondrites, final chondrule formation occurred under very different conditions.

## 11. Summary

By studying the composition of matrices in carbonaceous chondrites we have substantiated our earlier claims that chondrules and matrix in carbonaceous chondrites are chemically complementary and must have formed from single, chondritic nebular reservoirs. We have shown that the complementarity cannot be explained by alteration processes on the parent bodies of carbonaceous chondrites. Many of the prevailing chondrule formation models are incompatible with this finding. The single reservoir for chondrule and matrix formation was open for exchange with gases and the addition of minor components carrying stable isotope anomalies and presolar grains.

## Acknowledgements

We thank John Spratt for his competent assistance with electron microprobe analyses at the Natural History Museum in London. The paper benefited from comments by N.A. Abreu and the editor (T.E.). H.P. is grateful for support by the European Research Council (ERC) Advanced Grant "ACCRETE" (contract number 290568).

## Appendix A. Supplementary material

Supplementary material related to this article can be found on-line at http://dx.doi.org/10.1016/j.epsl.2014.11.033.

## References

Abreu, N.A., Brearley, A.J., 2010. Early solar system processes recorded in the matrices of two highly pristine CR3 carbonaceous chondrites, MET 00426 and QUE 99177. Geochim. Cosmochim. Acta 74, 1146–1171.

Ahrens, L.H., 1965. Observations on the Fe-Si-Mg relationship in chondrites. Geochim. Cosmochim. Acta 29, 801-806.

Palme, Hezel, Ebel (2015, *EPSL)*

Alexander, C.M.O'D, 1994. Trace element distributions within ordinary chondrite chondrules: implications for chondrule formation conditions and precursors. Geochim. Cosmochim. Acta 58, 3451–3467.

Asphaug, E., Jutzi, M., Movshovitz, M., 2011. Chondrule formation during planetesimal accretion. Earth Planet. Sci. Lett. 308, 369-379.

Berlin, J., 2009. Mineralogy and bulk chemistry of chondrules and matrix in petro-logic type 3 chondrules: implications for early solar system processes. Ph.D. thesis. University of New Mexico.

Bischoff, A., Palme, H., Ash, R.D., Clayton, R.N., Schultz, L., Herpers, U., Stöffler, D., Grady, M.M., Pillinger, C.T., Spettel, B., Weber, H., Grund, T., Endreß, M. , Weber, D., 1993. Paired Renazzo-type (CR) carbonaceous chondrites from the Sahara. Geochim. Cosmochim. Acta 57, 1587-1603.

Bischoff, A., Vogel, N., Roszjar, J., 2011. The Rumuruti chondrite group. Chemie der Erde 71, 101–133.

Bland, P.A., Alard, O., Benedix, G.K., Kearsley, A.T., Menzies, O.N., Watt, L.E., Rogers, N.W., 2005. Volatile fractionation in the early solar system and chondrule/matrix complementarity. Proc. Natl. Acad. Sci. USA 102, 13755–13760.

Brearley, A.J., 1993. Matrix and fine-grained rims in the unequilibrated CO3 chondrite, ALHA77307: origins and evidence for diverse, primitive nebular dust components. Geochim.Cosmochim. Acta 57, 1521–1550.

Brearley, A.J., 2003. Nebular versus parent-body processing. In: Holland, H.D., Turekian, K.K. (Eds.), Treatise on Geochemistry, vol. 5. Elsevier, Oxford, pp.225–272.

Ciesla, F., 2005. Chondrule-forming processes – An overview. In: Chondrites and the Protoplanetary Disk (eds Krot A.N. et al.) 811-820, Astronomical Society of the Pacific, San Francisco.

Clarke, R. S., Jarosewich, E., Mason, B., Nelen, J., Gomez, M. and Hyde, J. R., 1971. The Allende, Mexico, Meteorite Shower. Smithson. Contrib. Earth Sci. 5, 1-53.

Clayton, R.N., Onuma, N., Ikeda, Y., Mayeda, T.K., Hutcheon, I.D., Olsen, E.J.,Molini-Velsko, C., 1983. Oxygen isotopic compositions of chondrules in Allende and ordinary chondrites. In Chondrules and their Origins (ed. E.A. King). Lunar and Planetary Institute, Houston, pp. 37–43.

Connolly, H.C., Love, S.G., 1998. The formation of chondrules. Petrologic tests of the shock wave model. Science 280, 62-67.

Desch, S.J., Ciesla, F.J., Hood, L.L., Nakamoto, T., 2005. Heating of chondritic material s in solar nebula shocks. In Chondrites and the protoplanetary disk, (eds. Krot A. N., Scott E. R. D., and Reipurth B. San Francisco: Astronomical Society of the Pacific. pp. 849 -872.

Ebel, D.S., Weisberg, M.K., Hertz, J., and Campbell, 2008. A.J. Shape, metal abundance, chemistry and origin of chondrules in the Renazzo (CR) chondrite. Meteorit. Planet. Sci. 43, 1725-1740.

Ebel, D.S., Leftwich, K., Brunner, C.E., Weisberg, M.K., 2009. Abundance and size distribution of inclusions in CV3 chondrites by X-ray image analysis. In: 40th Lunar and Planetary Science Conference. #2065 (abstr.).

Gerber, M., 2012. Ph.D. thesis. Universität Münster, Germany.

Grossman, J.N., Brearley, A.J., 2005. The onset of metamorphism in ordinary and carbonaceous chondrites. Meteorit. Planet. Sci. 40, 87–122.

Grossman, J.N., Wasson, J.T., 1983. The compositions of chondrules in unequilibrated chondrites: An evaluation of models for the formation of chondrules and their precursor materials. In Chondrules and Their Origins (ed. E. A. King), pp. 88-121. Lunar and Planetary Institute, Houston, TX.

Grossman, J.N., Wasson, J.T., 1985. The origin and history of the metal and sulfide components of chondrules. Geochim. Cosmochim. Acta 49, 925–939.

Grossman, J.N., Alexander, C.M.O'd., Wang, J., Brearley, A.J., 2002. Zoned chondrules in Semarkona: evidence for high-and low-temperature processing. Meteorit. Planet. Sci. 37, 49–73.

Hewins, R., 1997. Chondrules. Annu. Rev. Earth Planet. Sci. 25, 61–83.

Hezel, D.C., Kießwetter, R., 2010. Quantifying the error of 2D bulk chondrule analyses using a computer model to simulate chondrules (SIMCHON). Meteorit. Planet. Sci. 45, 555–571.

Hezel, D.C., Palme, H. 2007. The conditions of chondrule formation, Part I: Closed system. Geochim. Cosmochim. Acta 71, 4092-4107.

Hezel, D.C., Palme, H., 2008. Constraints for chondrule formation from Ca–Al distribution in carbonaceous chondrites. Earth Planet. Sci. Lett. 265, 716–725.

Hezel, D.C. and Palme, H., 2010. The chemical relationship between chondrules and matrix and the chondrule matrix complementarity. Earth Planet. Sci. Lett. 294, 85-93.

Hezel, D.C., Palme, H., Nasdala, L., Brenker, F.E., 2006. Origin of SiO2-rich components in ordinary chondrites. Geochim. Cosmochim. Acta70, 1548–1564.

Hezel, D.C., Russell, S.S., Ross, A.J., Kearsley, A.T., 2008. Modal abundances of CAIs: implications for bulk chondrite element abundances and fractionations. Meteorit. Planet. Sci. 43, 1879–1894.

Hezel, D.C., Elangovan, P., Viehmann, S., Howard, L., Abel, R.L., Armstrong, R., 2013. Visualisation and quantification of CV chondrite petrography using micro-tomography. Geochim. Cosmochim. Acta 116, 33–40.

Hua, X., Adam, J., Palme, H., El Goresy, A., 1988. Fayalite-rich rims, veins, and ha-los around and in forsteritic olivines in CAIs and chondrules in carbonaceous chondrites: types, compositional profiles, and constraints of their formation. Geochim. Cosmochim. Acta 52, 1389–1408.

Huss, G.R., Keil, K. and Taylor, G.J., 1981. The matrices of unequilibrated ordinary chondrites: implications for the origin and history of chondrites. Geochim. Cosmochim. Acta 45, 33-51).

Huss, G.R., Meshik, A.P., Smith, J.B., Hohenberg, C.M., 2003. Presolar diamond, silicon carbide, and graphite in carbonaceous chondrites: implications for thermal processing in the solar nebula. Geochim. Cosmochim. Acta 67, 4823-4848.

Huss, G. R., Alexander, C. M. O., Palme, H., Bland, P. A., and Wasson, J. T., 2005. Genetic relationship between chondrules, fine-grained rims, and interchondrule matrix. In Chondrites and the protoplanetary disk, (eds. Krot A. N., Scott E. R. D., and Reipurth B. San Francisco: Astronomical Society of the Pacific. pp. 701-731.

Palme, Hezel, Ebel (2015, *EPSL)*

Ikeda, Y., Kimura, M., 1995. Anhydrous alteration of Allende chondrules in the solar nebula I: description and alteration of chondrules with known oxygen-isotopic compositions. Proc. NIPR Symp. Antarct. Meteor. 8, 97–122.

Jarosewich, E., 1990. Chemical analyses of meteorites: A compilation of stony and iron meteorites. Meteoritics 25, 323-337.

Jarosewich, E., Clarke, R.S., Barrows, J.N., 1987. The Allende meteorite reference sample. Smithson. Contrib. Earth Sci. 27, 1–12.

Jones, R.H., Schilk, A.J., 2009. Chemistry, petrology and bulk oxygen isotope compositions of chondrules from the Mokoia CV3 carbonaceous chondrite. Geochim. Cosmochim. Acta 73, 5854–5883.

Jones, R.H., Leshin, L.A., Guan, Y., Sharp, Z.D., Durakiewicz, T. and Schilk, A.J., 2004. Oxygen isotope heterogeneity in chondrules from the Mokoia CV3 carbonaceous chondrite. Geochim. Cosmochim. Acta 68, 3423–3438.

Jones, R H., Grossman, J.N., Rubin, A.E., 2005. Chemical, mineralogical and isotopic properties of chondrules: Clues to their origins. In: Chondrites and the protoplanetary disk (eds Krot A.N. et al.) 251–285, Astronomical Society of the Pacific, San Francisco.

Kallemeyn, G.W., Wasson, J.T.,1981. The compositional classification of chondrites – I. The carbonaceous chondrite groups. Geochim. Cosmochim. Acta 45, 1217-1230.

Kallemeyn, G.W., Rubin, A.E., Wasson, J.T., 1994. The compositional classification of chondrites: VI. The CR carbonaceous chondrite group. Geochim. Cosmochim. Acta 58, 2873–2888.

Kimura, M., Ikeda, Y., 1995. Anhydrous alteration of Allende chondrules in the solar nebula II: alkali–Ca exchange reactions and formation of nepheline, sodalite and Ca-rich phases in chondrules. Proc. NIPR Symp. Antarct. Meteor. 8, 123–138.

Kimura, M., Ikeda, Y., 1998. Hydrous and anhydrous alteration of chondrules in Kaba and Mokoia CV chondrites. Meteorit. Planet. Sci. 33, 1139-1146.

Kita, N.T., Tenner, T.J., Ushikubo, T., Nakashima, D., Bischoff, A., 2013. Primitive chondrules in a highly unequilibrated clast in NWA 753 R chondrite. Lunar and Planetary Science 44, (abstract) 1784.

Klerner, S., 2001. Materie im frühen Sonnensystem: Die Entstehung von Chondren, Matrix, und refraktären Forsteriten. PhD Thesis, Universität zu Köln.

Klerner, S., Palme, H., 1999. Origin of Chondrules and Matrix in Carbonaceous Chondrites. Lunar and Planetary Science Conference 30, #1272, Lunar and Planetary Institute, Houston.

Krot, A.N., Scott, E.R.D., Zolensky, M.E., 1995. Mineralogic and chemical variations among CV3 chondrites and their components: nebular and asteroidal processing. Meteoritics 30, 748–775.

Krot, A.N., Petaev, M.I., Scott, E.R.D., Choi, B.-G., Zolensky, M.E., Keil, K., 1998. Progressive alteration in CV3 chondrites: more evidence for asteroidal alteration. Meteorit. Planet. Sci. 33, 1065–1085.

Lodders, K., Palme, H., Gail, H.P., 2009. In: J. E. Trümper (Ed.), Abundances of the elements in the solar system. Landolt-Börnstein, New Series, Vol. VI/4B, Springer, Berlin, pp. 560–598.

Palme, Hezel, Ebel (2015, *EPSL*)

Lugmair, G.W., Shukolyukov, A., 2001. Early solar system events and timescales. Meteorit. Planet. Sci. 36, 1017-1026.

MacPherson, G.J.M., Krot, A.N., 2014. The formation of Ca-, Fe-rich silicates in re-duced and oxidized CV chondrites: the roles of impact-modified porosity and permeability, and heterogeneous distribution of water ices. Meteorit. Planet. Sci. 49, 1250–1270.

McNally, C.P., Hubbard, A., Mac Low, M-M., Ebel, D.S., and D'Alessio, P., 2013. Mineral processing by short circuits in protoplanetary disks. Astrophys. J. 767, L2-L7.

Misawa, K., Nakamura, N., 1988. Demonstration of REE fractionation among individual chondrules from the Allende (CV3) chondrite. Geochim. Cosmochim. Acta 52, 1699-1710

Murakami, T., Ikeda, Y.,1994. Petrology and mineralogy of the Yamato-86751 CV3 chondrite. Meteorit. Planet. Sci. 29, 397–409.

Palme, H., Spettel, B., Hezel, D.C., 2014a. Siderophile elements in chondrules of CV chondrites. Chem. Erde -Geochem. 74, 507–514.

Palme, H., Lodders, K., Jones, A., 2014b. Solar system abundances of the elements. In: Holland, H.D., Turekian, K.K. (Eds.), Treatise on Geochemistry, vol. 2. Second edition. Elsevier, Oxford, pp.15–36.

Pouchou, J.L., Pichoir, F., 1991. Quantitative analysis of homogeneous or stratified mi-crovolumes applying the model "PAP". In: Heinrich, K.F.J., Newbury, D.E. (Eds.), Electron Probe Quantification. Plenum, New York, pp.31–75.

Rocholl, A., Jochum, K.P., 1993. Th, U and other trace elements in carbonaceous chondrites: implications for the terrestrial and solar-system Th/U ratios. Earth Planet. Sci. Lett. 117, 265–278.

Rubin, A.E., Wasson, J.T., 1987. Chondritdes, matrix and coarse-grained chondrule rims in the Allende meteorite: Origin. Interrelationships and possibIe precursor components. Geochim. Cosmochim. Acta 51, 1923-1937.

Rudraswami, N.G., Ushikubo, N., Nakashima, D., Kita, N.T., 2011. Oxygen isotope systematics of chondrules in the Allende CV3 chondrite: High precision ion microprobe studies. Geochim. Cosmochim. Acta 75, 7596–7611.

Sanders, I.S., Scott, E.R.D., 2012. The origin of chondrules and chondrites: debris from low-velocity impacts between molten planetesimals? Meteorit. Planet. Sci. 47, 2170-2192.

Savage, K.B., Sembach, D.R., 1996.Interstellar abundances from absorption-line observations with the Hubble Space Telescope, Ann. Rev. Astron. & Astrophys. 34, 279-329.

Schilk, A.J., 1991. Ph.D. thesis. Oregon State University.

Shu, F. H., Shang H., Gounelle M., Glassgold A. E., and Lee, T., 2001. The origin of chondrules and refractory inclusions in chondritic meteorites. Astrophys. J. 548, 1029–1050.

Stracke, A., Palme, H., Gellissen, M., Münker, C., Kleine, T., Birbaum, K., Günther, D., Bourdon, B., Zipfel J., 2012. Refractory element fractionation in the Allende meteorite: Implications for solar nebula condensation and the chondritic composition of planetary bodies. Geochim. Cosmochim. Acta 85, 114-141.

Wasson, J. T., Rubin, A. E., 2009. Composition of matrix in the CR chondrite LAP 02342. Geochim. Cosmochim. Acta 73, 1436–1460.

Wasson, J.T., Rubin, A.E., 2010a. Matrix and whole-rock fractionations in the Acfer 094 type 3.0 ungrouped carbonaceous chondrite. Meteorit. Planet. Sci. 45, 73–90.

Wasson, J.T., Rubin, A.E., 2010b. Metal in CR chondrites. Geochim. Cosmochim. Acta 74, 2212–2230.

Weisberg M. K., McCoy T. J., Krot, A.N., 2006. Systematics and evaluation of meteorite classification. In: Meteorites and the Early Solar System II, (eds. D. Lauretta D., H.Y. McSween H.Y. Jr.), University of Arizona, Tucson. pp. 19-52

Wiik, H.B., (1956) The chemical composition of some stony meteorites. Geochim. Cosmochim. Acta 9, 279-289.

Wolf, D. and Palme, H., 2001. The solar system abundances of phosphorus and titanium and the nebular volatility of phosphorus Meteorit. Planet. Sci. 36, 559-571.

Wood, J. A. (1963). On the origin of chondrules and chondrites. Icarus 2,152–180.

Wood, J.A., 1985. Meteoritic constraints on processes in the solar nebula. In: Black, D.C., Matthews, M.S. (Eds.), Protostars and Planets, vol. II. Univ. of Arizona Press, Tucson, pp.687–702.

Zolensky, M., Barrett, R., Browning, R., 1993. Mineralogy and composition of matrix and chondrule rims in carbonaceous chondrites. Geochim. Cosmochim. Acta 57, 3123–3148.

Zook, H. A., 1981. On a new model for the generation of chondrules (abstract). Lunar and Planetary Science Conference 12, 1242-1244.

Palme, Hezel, Ebel (2015, *EPSL*)

**Supplementary Online Material:** Additional electron microprobe data of (**Table S1**) matrix in the CV3-chondrite Mokoia and of (**Table S2**) chondrules in the CR2 meteorite Renazzo. The Mokoia matrix data are used in Fig. 5 and the Renazzo chondrule data in Fig. 7. The latter data are reproduced from Klerner (2001).

**Table S1**: Matrix data for Mokoia. Measurements were performed with the Cameca SX100 electron microprobe at the Natural History Museum, London. The electron beam was defocused to a diameter of 20 µm.

| No. | $SiO_2$ | $TiO_2$ | $Al_2O_3$ | $Cr_2O_3$ | FeO | MnO | NiO | MgO | CaO | $Na_2O$ | Total |
|---|---|---|---|---|---|---|---|---|---|---|---|
| 1 | 28.37 | 0.11 | 2.96 | 0.35 | 38.31 | 0.23 | 1.24 | 18.69 | 0.11 | 0.14 | 90.60 |
| 2 | 29.56 | 0.11 | 1.43 | 0.39 | 38.58 | 0.21 | 0.90 | 20.17 | 0.29 | 0.13 | 91.86 |
| 3 | 30.44 | 0.05 | 1.24 | 0.40 | 35.64 | 0.20 | 0.25 | 21.42 | 0.16 | 0.18 | 90.05 |
| 4 | 31.01 | 0.02 | 0.33 | 0.12 | 39.15 | 0.25 | 0.19 | 20.27 | 0.10 | 0.08 | 91.56 |
| 5 | 28.30 | 0.05 | 1.19 | 0.13 | 40.11 | 0.24 | 0.67 | 19.01 | 0.09 | 0.59 | 90.46 |
| 6 | 29.17 | 0.05 | 1.39 | 0.25 | 38.65 | 0.22 | 0.64 | 19.40 | 0.15 | 0.11 | 90.11 |
| 7 | 27.12 | 0.08 | 2.51 | 0.41 | 38.29 | 0.24 | 1.60 | 18.71 | 0.15 | 0.14 | 89.36 |
| 8 | 29.62 | 0.09 | 1.51 | 0.49 | 34.96 | 0.21 | 0.76 | 21.44 | 0.26 | 0.13 | 89.55 |
| 9 | 29.99 | 0.13 | 1.35 | 0.35 | 33.54 | 0.19 | 0.62 | 22.11 | 0.14 | 0.28 | 88.76 |
| 10 | 29.28 | 0.06 | 1.59 | 0.39 | 37.51 | 0.23 | 1.22 | 20.04 | 0.09 | 0.14 | 90.67 |
| 11 | 28.60 | 0.07 | 2.24 | 0.91 | 38.10 | 0.21 | 1.67 | 17.59 | 1.35 | 0.13 | 91.01 |
| 12 | 28.45 | 0.06 | 2.02 | 0.34 | 38.87 | 0.25 | 0.93 | 18.76 | 0.12 | 0.12 | 90.01 |
| 13 | 28.30 | 0.06 | 0.66 | 0.26 | 38.92 | 0.25 | 1.57 | 18.88 | 0.16 | 0.12 | 89.28 |
| 14 | 29.48 | 0.03 | 2.04 | 0.26 | 39.92 | 0.23 | 1.87 | 20.12 | 0.10 | 0.11 | 94.28 |
| 15 | 25.64 | 0.05 | 3.27 | 0.40 | 38.26 | 0.18 | 2.52 | 18.07 | 0.24 | 0.16 | 88.97 |
| 16 | 28.52 | 0.06 | 1.34 | 0.26 | 37.68 | 0.22 | 2.27 | 19.96 | 0.07 | 0.15 | 90.67 |
| 17 | 30.92 | 0.08 | 1.21 | 0.38 | 33.58 | 0.19 | 0.72 | 18.48 | 2.56 | 0.18 | 88.37 |
| 18 | 29.83 | 0.08 | 1.54 | 0.29 | 32.42 | 0.21 | 1.45 | 19.04 | 2.29 | 0.18 | 87.44 |
| 19 | 42.85 | 0.01 | 0.38 | 0.12 | 13.64 | 0.94 | 0.27 | 8.11 | 22.61 | 0.14 | 89.12 |
| 20 | 30.79 | 0.10 | 1.95 | 1.17 | 33.25 | 0.21 | 0.84 | 21.72 | 0.82 | 0.39 | 91.31 |
| 21 | 28.30 | 0.11 | 1.23 | 0.36 | 37.96 | 0.23 | 0.41 | 19.72 | 0.14 | 0.10 | 88.62 |
| 22 | 30.53 | 0.07 | 2.70 | 0.28 | 34.92 | 0.21 | 1.27 | 16.52 | 3.38 | 0.26 | 90.24 |
| 23 | 26.88 | 0.07 | 1.75 | 0.59 | 38.05 | 0.24 | 1.68 | 18.04 | 0.11 | 0.14 | 87.65 |
| 24 | 29.31 | 0.10 | 0.97 | 0.30 | 37.93 | 0.24 | 1.72 | 20.33 | 0.07 | 0.12 | 91.20 |
| 25 | 29.27 | 0.05 | 1.85 | 0.38 | 41.81 | 0.27 | 0.65 | 18.92 | 0.16 | 0.12 | 93.56 |
| 26 | 28.61 | 0.10 | 1.30 | 0.33 | 40.80 | 0.26 | 0.86 | 18.03 | 0.17 | 0.23 | 90.80 |
| 27 | 27.89 | 0.11 | 2.95 | 0.43 | 34.62 | 0.16 | 2.11 | 21.51 | 0.08 | 0.21 | 90.21 |
| 28 | 27.37 | 0.08 | 2.33 | 0.32 | 41.29 | 0.25 | 0.81 | 17.30 | 0.10 | 0.13 | 90.06 |
| 29 | 28.45 | 0.10 | 2.88 | 0.48 | 36.04 | 0.22 | 1.22 | 20.06 | 0.10 | 0.27 | 89.94 |
| 30 | 34.53 | 0.22 | 2.22 | 0.35 | 25.07 | 0.16 | 0.73 | 27.68 | 0.12 | 0.59 | 91.79 |
| 31 | 32.32 | 0.18 | 1.75 | 0.32 | 26.09 | 0.15 | 1.21 | 26.42 | 0.12 | 0.53 | 89.21 |
| 32 | 31.43 | 0.16 | 2.33 | 0.30 | 27.66 | 0.16 | 0.87 | 26.00 | 0.09 | 0.44 | 89.55 |
| 33 | 37.39 | 0.27 | 4.11 | 0.16 | 21.62 | 0.16 | 0.08 | 31.48 | 1.28 | 1.14 | 97.82 |

| | | | | | | | | | | |
|---|---|---|---|---|---|---|---|---|---|---|
| 34 | 29.33 | 0.13 | 1.71 | 0.29 | 39.33 | 0.20 | 1.05 | 20.78 | 0.11 | 0.13 | 93.15 |
| 35 | 28.20 | 0.08 | 2.51 | 0.34 | 40.69 | 0.26 | 0.71 | 18.38 | 0.15 | 0.25 | 91.67 |
| 36 | 26.04 | 0.04 | 2.22 | 0.30 | 40.83 | 0.23 | 3.05 | 17.09 | 0.34 | 0.12 | 90.41 |
| 37 | 23.47 | 0.08 | 5.07 | 0.26 | 41.23 | 0.21 | 2.23 | 16.02 | 0.32 | 0.14 | 89.19 |
| 38 | 29.88 | 0.09 | 1.97 | 0.26 | 33.97 | 0.20 | 1.34 | 22.72 | 0.07 | 0.17 | 90.78 |
| 39 | 30.18 | 0.04 | 1.11 | 0.22 | 38.17 | 0.25 | 1.02 | 20.72 | 0.09 | 0.12 | 92.01 |
| 40 | 14.89 | 0.07 | 10.13 | 0.23 | 42.15 | 0.10 | 10.35 | 12.47 | 0.00 | 0.16 | 91.03 |
| 41 | 21.66 | 0.07 | 3.36 | 0.94 | 40.84 | 0.21 | 5.44 | 14.76 | 0.11 | 0.12 | 87.77 |
| 42 | 26.47 | 0.08 | 2.81 | 0.32 | 39.92 | 0.20 | 2.21 | 17.00 | 0.08 | 0.13 | 89.37 |
| 43 | 25.38 | 0.07 | 3.07 | 0.26 | 40.08 | 0.22 | 1.36 | 17.19 | 0.07 | 0.14 | 87.95 |
| 44 | 26.20 | 0.06 | 3.15 | 0.59 | 39.73 | 0.23 | 2.43 | 16.83 | 0.57 | 0.17 | 90.10 |
| 45 | 28.43 | 0.08 | 1.54 | 0.29 | 39.72 | 0.22 | 1.31 | 19.32 | 0.08 | 0.11 | 91.24 |
| 46 | 28.37 | 0.06 | 1.69 | 0.54 | 37.55 | 0.21 | 0.98 | 19.03 | 0.16 | 0.13 | 88.83 |
| 47 | 26.56 | 0.08 | 2.07 | 0.41 | 43.82 | 0.24 | 1.04 | 17.03 | 0.37 | 0.12 | 91.83 |
| 48 | 32.68 | 0.02 | 0.86 | 0.20 | 37.41 | 0.31 | 0.58 | 17.28 | 2.56 | 0.15 | 92.14 |
| 49 | 33.52 | 0.01 | 0.12 | 0.13 | 42.09 | 0.28 | 0.10 | 21.85 | 0.12 | 0.02 | 98.30 |
| 50 | 28.01 | 0.08 | 2.93 | 0.37 | 41.75 | 0.26 | 1.09 | 17.23 | 0.10 | 0.23 | 92.14 |
| 51 | 27.34 | 0.05 | 2.17 | 0.44 | 42.22 | 0.23 | 0.75 | 16.54 | 0.35 | 0.48 | 90.67 |
| 52 | 27.45 | 0.07 | 1.39 | 0.40 | 42.21 | 0.24 | 2.02 | 16.35 | 0.31 | 0.39 | 90.99 |
| 53 | 27.51 | 0.07 | 2.47 | 0.74 | 34.91 | 0.27 | 4.22 | 13.83 | 4.52 | 0.52 | 89.29 |
| 54 | 30.47 | 0.04 | 0.35 | 0.23 | 41.51 | 0.27 | 0.79 | 18.49 | 0.09 | 0.45 | 92.80 |
| 55 | 28.68 | 0.07 | 1.17 | 0.37 | 41.56 | 0.25 | 0.93 | 17.94 | 0.30 | 0.49 | 91.85 |
| 56 | 26.44 | 0.06 | 1.71 | 0.30 | 43.17 | 0.23 | 2.43 | 16.45 | 0.09 | 0.42 | 91.48 |
| 57 | 29.90 | 0.07 | 0.94 | 0.34 | 40.69 | 0.29 | 0.81 | 18.92 | 0.09 | 0.52 | 92.67 |
| 58 | 28.92 | 0.04 | 2.51 | 0.28 | 41.66 | 0.27 | 0.86 | 18.25 | 0.10 | 0.46 | 93.47 |
| 59 | 24.68 | 0.05 | 1.77 | 0.28 | 41.01 | 0.23 | 3.48 | 16.08 | 0.64 | 0.39 | 88.82 |
| 60 | 31.00 | 0.07 | 2.36 | 0.41 | 36.61 | 0.24 | 0.75 | 17.85 | 2.66 | 0.53 | 92.60 |
| 61 | 28.65 | 0.04 | 1.78 | 0.32 | 40.82 | 0.26 | 1.34 | 18.53 | 0.18 | 0.45 | 92.50 |
| 62 | 30.41 | 0.10 | 1.49 | 0.40 | 31.50 | 0.18 | 1.37 | 23.60 | 0.10 | 0.72 | 90.01 |
| 63 | 43.59 | 0.03 | 1.45 | 0.09 | 15.61 | 0.14 | 0.67 | 8.49 | 20.71 | 0.79 | 91.64 |
| 64 | 29.66 | 0.07 | 1.20 | 0.43 | 38.32 | 0.22 | 1.34 | 20.10 | 0.11 | 0.63 | 92.20 |
| 65 | 28.23 | 0.06 | 1.83 | 0.49 | 40.48 | 0.26 | 1.62 | 18.41 | 0.22 | 0.36 | 92.10 |
| 66 | 28.47 | 0.09 | 1.06 | 0.35 | 40.07 | 0.23 | 0.48 | 20.39 | 0.06 | 0.58 | 91.84 |
| 67 | 27.24 | 0.09 | 1.85 | 0.40 | 40.99 | 0.21 | 1.32 | 18.02 | 0.06 | 0.59 | 90.90 |
| 68 | 31.05 | 0.06 | 1.11 | 0.31 | 36.45 | 0.25 | 0.69 | 22.18 | 0.05 | 0.63 | 92.88 |
| 69 | 28.98 | 0.06 | 1.56 | 0.25 | 38.65 | 0.24 | 1.02 | 19.54 | 0.08 | 0.65 | 91.13 |
| 70 | 30.17 | 0.09 | 2.24 | 0.43 | 37.79 | 0.23 | 1.00 | 21.06 | 0.09 | 0.40 | 93.62 |
| 71 | 28.01 | 0.08 | 2.01 | 0.68 | 39.67 | 0.24 | 1.15 | 18.40 | 0.10 | 0.56 | 91.04 |
| 72 | 31.54 | 0.02 | 0.24 | 0.16 | 36.01 | 0.24 | 0.03 | 21.64 | 0.21 | 0.87 | 91.04 |
| 73 | 26.36 | 0.07 | 3.04 | 0.63 | 38.00 | 0.20 | 2.00 | 18.73 | 0.10 | 0.60 | 89.84 |
| 74 | 29.77 | 0.08 | 0.94 | 0.26 | 38.78 | 0.20 | 1.67 | 20.43 | 0.08 | 0.52 | 92.88 |
| 75 | 36.34 | 0.06 | 0.99 | 0.30 | 28.73 | 0.19 | 0.26 | 13.21 | 10.77 | 0.53 | 91.46 |
| 76 | 30.77 | 0.05 | 0.29 | 0.21 | 40.01 | 0.26 | 0.30 | 19.60 | 0.09 | 0.30 | 91.95 |
| 77 | 28.47 | 0.03 | 1.02 | 0.29 | 38.54 | 0.24 | 0.50 | 18.12 | 0.11 | 0.32 | 87.73 |

|  |  |  |  |  |  |  |  |  |  |  |
|---|---|---|---|---|---|---|---|---|---|---|
| 78  | 21.60 | 0.04 | 4.08  | 0.51 | 43.02 | 0.19 | 6.72  | 15.01 | 0.13  | 0.21 | 91.86 |
| 79  | 41.48 | 0.01 | 0.90  | 0.11 | 13.82 | 0.66 | 0.24  | 6.66  | 19.90 | 0.58 | 84.39 |
| 80  | 29.80 | 0.09 | 1.00  | 0.36 | 39.92 | 0.26 | 0.48  | 19.02 | 0.16  | 0.25 | 91.46 |
| 81  | 28.95 | 0.08 | 1.21  | 0.48 | 39.23 | 0.27 | 0.77  | 18.74 | 0.12  | 0.23 | 90.20 |
| 82  | 31.35 | 0.07 | 0.76  | 0.27 | 35.84 | 0.25 | 0.34  | 18.57 | 1.95  | 0.31 | 89.78 |
| 83  | 26.69 | 0.06 | 2.64  | 0.34 | 36.62 | 0.19 | 2.68  | 19.41 | 0.12  | 0.33 | 89.25 |
| 84  | 27.38 | 0.09 | 2.77  | 0.49 | 37.49 | 0.23 | 0.75  | 18.61 | 0.17  | 0.30 | 88.39 |
| 85  | 32.53 | 0.08 | 1.64  | 0.51 | 30.93 | 0.21 | 1.00  | 15.15 | 6.64  | 0.30 | 89.11 |
| 86  | 27.90 | 0.06 | 2.18  | 0.39 | 38.36 | 0.25 | 0.95  | 17.50 | 0.40  | 0.25 | 88.35 |
| 87  | 29.43 | 0.09 | 1.20  | 0.41 | 38.12 | 0.25 | 1.37  | 17.33 | 1.24  | 0.27 | 89.84 |
| 88  | 29.82 | 0.06 | 0.90  | 0.30 | 40.91 | 0.27 | 0.84  | 18.75 | 0.29  | 0.25 | 92.50 |
| 89  | 24.99 | 0.06 | 2.61  | 0.26 | 40.74 | 0.21 | 2.31  | 16.22 | 0.09  | 0.24 | 87.94 |
| 90  | 30.71 | 0.11 | 1.17  | 0.37 | 36.12 | 0.23 | 0.96  | 21.44 | 0.18  | 0.33 | 91.72 |
| 91  | 27.57 | 0.06 | 2.11  | 0.44 | 40.41 | 0.24 | 1.54  | 18.10 | 0.12  | 0.38 | 91.13 |
| 92  | 32.10 | 0.02 | 0.49  | 0.29 | 43.49 | 0.30 | 0.49  | 19.56 | 0.10  | 0.25 | 97.19 |
| 93  | 30.25 | 0.07 | 0.57  | 0.18 | 41.28 | 0.25 | 0.57  | 19.28 | 0.10  | 0.41 | 93.05 |
| 94  | 29.92 | 0.03 | 0.76  | 0.26 | 41.23 | 0.27 | 0.69  | 18.80 | 0.06  | 0.36 | 92.49 |
| 95  | 28.36 | 0.06 | 2.07  | 0.44 | 38.76 | 0.23 | 1.08  | 19.24 | 0.09  | 0.39 | 90.84 |
| 96  | 24.30 | 0.06 | 4.68  | 0.38 | 42.25 | 0.24 | 3.00  | 15.45 | 0.09  | 0.47 | 91.12 |
| 97  | 25.80 | 0.05 | 3.56  | 0.23 | 39.76 | 0.20 | 2.11  | 17.37 | 0.14  | 0.65 | 90.04 |
| 98  | 30.19 | 0.11 | 2.84  | 0.51 | 29.73 | 0.18 | 1.71  | 22.60 | 0.09  | 0.95 | 89.14 |
| 99  | 30.13 | 0.07 | 1.31  | 0.27 | 32.52 | 0.18 | 0.79  | 23.09 | 0.08  | 0.87 | 89.43 |
| 100 | 29.78 | 0.05 | 1.22  | 0.20 | 39.85 | 0.27 | 0.49  | 19.10 | 0.10  | 0.57 | 91.72 |
| 101 | 29.37 | 0.05 | 1.18  | 0.21 | 40.84 | 0.26 | 0.85  | 18.89 | 0.13  | 0.53 | 92.42 |
| 102 | 14.05 | 4.79 | 42.51 | 0.03 | 2.68  | 0.02 | -0.01 | 7.05  | 26.67 | 0.30 | 98.09 |
| 103 | 1.01  | 1.64 | 67.24 | 0.14 | 1.98  | 0.03 | 0.02  | 18.78 | 2.13  | 0.23 | 93.20 |
| 104 | 19.02 | 8.10 | 31.62 | 0.02 | 0.08  | 0.00 | -0.01 | 0.71  | 40.98 | 0.15 | 100.67 |
| 105 | 35.59 | 0.17 | 39.17 | 0.01 | 1.63  | 0.02 | 0.01  | 2.60  | 16.17 | 1.24 | 96.62 |
| 106 | 28.05 | 0.08 | 3.23  | 0.30 | 35.05 | 0.20 | 1.30  | 19.49 | 0.09  | 0.99 | 88.92 |
| 107 | 26.29 | 0.05 | 3.18  | 0.26 | 37.91 | 0.20 | 2.02  | 19.27 | 0.08  | 0.48 | 89.95 |
| 108 | 28.76 | 0.06 | 1.57  | 0.28 | 39.32 | 0.24 | 1.69  | 18.90 | 0.12  | 0.37 | 91.49 |
| 109 | 29.15 | 0.07 | 1.61  | 0.30 | 42.81 | 0.28 | 0.76  | 18.73 | 0.17  | 0.28 | 94.29 |
| 110 | 30.40 | 0.06 | 0.71  | 0.31 | 40.38 | 0.26 | 1.11  | 18.85 | 0.40  | 0.35 | 92.93 |
| 111 | 31.05 | 0.07 | 1.66  | 0.26 | 37.27 | 0.23 | 0.59  | 20.39 | 0.66  | 0.77 | 93.09 |
| 112 | 25.19 | 0.04 | 3.91  | 0.39 | 37.66 | 0.19 | 2.02  | 17.14 | 0.13  | 0.32 | 87.15 |
| 113 | 28.54 | 0.03 | 1.53  | 0.40 | 39.11 | 0.23 | 0.99  | 19.03 | 0.14  | 0.37 | 90.50 |
| 114 | 22.41 | 0.08 | 6.10  | 0.61 | 42.10 | 0.17 | 2.59  | 15.05 | 0.14  | 0.43 | 89.89 |
| 115 | 30.17 | 0.07 | 0.87  | 0.35 | 37.82 | 0.23 | 0.44  | 20.57 | 0.19  | 0.66 | 91.48 |
| 116 | 25.39 | 0.07 | 2.59  | 0.45 | 40.23 | 0.18 | 3.23  | 17.42 | 0.40  | 0.96 | 91.15 |
| 117 | 38.00 | 0.05 | 1.32  | 0.20 | 18.80 | 0.37 | 0.45  | 13.49 | 13.44 | 1.31 | 87.49 |
| 118 | 27.12 | 0.09 | 3.18  | 0.61 | 38.15 | 0.22 | 1.72  | 18.28 | 0.63  | 0.54 | 90.71 |
| 119 | 27.76 | 0.04 | 1.80  | 0.29 | 38.22 | 0.20 | 2.48  | 19.72 | 0.13  | 0.38 | 91.20 |
| 120 | 29.16 | 0.06 | 1.65  | 0.40 | 39.92 | 0.27 | 1.25  | 18.96 | 0.14  | 0.44 | 92.39 |
| 121 | 29.70 | 0.08 | 1.09  | 0.62 | 39.66 | 0.29 | 0.79  | 19.76 | 0.14  | 0.43 | 92.67 |

| 122 | 29.59 | 0.10 | 1.82 | 0.42 | 36.22 | 0.20 | 1.78 | 21.02 | 0.12 | 0.74 | 92.20 |
|---|---|---|---|---|---|---|---|---|---|---|---|
| 123 | 28.81 | 0.05 | 1.16 | 0.38 | 40.04 | 0.23 | 0.98 | 18.44 | 0.10 | 0.39 | 90.68 |
| 124 | 27.96 | 0.07 | 2.55 | 0.25 | 41.30 | 0.22 | 1.12 | 17.45 | 0.54 | 0.61 | 92.19 |
| 125 | 28.04 | 0.07 | 2.65 | 0.41 | 37.77 | 0.21 | 1.23 | 18.61 | 0.13 | 1.53 | 90.78 |
| 126 | 28.65 | 0.09 | 1.24 | 0.40 | 40.68 | 0.25 | 0.72 | 17.99 | 0.13 | 0.45 | 90.71 |
| 127 | 27.49 | 0.10 | 1.47 | 0.42 | 41.25 | 0.24 | 1.15 | 17.55 | 0.12 | 0.48 | 90.40 |
| 128 | 32.43 | 0.04 | 0.67 | 0.59 | 40.85 | 0.25 | 0.24 | 21.63 | 0.13 | 0.26 | 97.19 |
| 129 | 31.63 | 0.04 | 0.87 | 0.25 | 37.21 | 0.23 | 0.53 | 18.02 | 2.70 | 0.37 | 91.98 |
| 130 | 29.17 | 0.07 | 1.33 | 0.35 | 38.96 | 0.25 | 0.66 | 18.51 | 0.12 | 0.49 | 90.04 |
| 131 | 30.11 | 0.09 | 0.79 | 0.39 | 34.19 | 0.20 | 1.08 | 22.83 | 0.05 | 0.52 | 90.37 |
| 132 | 30.82 | 0.09 | 1.71 | 0.43 | 34.00 | 0.20 | 1.31 | 22.01 | 0.08 | 0.59 | 91.36 |
| 133 | 28.39 | 0.06 | 3.54 | 0.23 | 38.85 | 0.19 | 0.53 | 19.77 | 0.07 | 0.45 | 92.20 |
| 134 | 26.53 | 0.05 | 1.77 | 0.25 | 40.10 | 0.22 | 3.84 | 18.03 | 0.06 | 0.47 | 91.59 |
| 135 | 32.41 | 0.06 | 1.03 | 0.27 | 34.08 | 0.21 | 1.15 | 15.20 | 5.28 | 0.48 | 90.29 |
| 136 | 30.92 | 0.05 | 0.69 | 0.20 | 39.81 | 0.25 | 0.58 | 20.37 | 0.08 | 0.36 | 93.41 |
| 137 | 21.11 | 0.05 | 3.20 | 0.28 | 44.70 | 0.17 | 5.57 | 13.97 | 0.09 | 0.44 | 89.94 |
| 138 | 42.36 | 0.06 | 1.26 | 0.19 | 17.39 | 0.54 | 0.43 | 11.16 | 16.67 | 0.75 | 90.90 |
| 139 | 27.78 | 0.09 | 2.36 | 0.27 | 42.39 | 0.22 | 1.35 | 18.16 | 0.10 | 0.50 | 93.37 |
| 140 | 23.62 | 0.02 | 0.22 | 0.11 | 18.38 | 0.15 | 0.00 | 19.65 | 0.26 | 1.20 | 63.72 |
| 141 | 29.92 | 0.06 | 1.11 | 0.32 | 38.70 | 0.23 | 0.86 | 20.63 | 0.10 | 0.41 | 92.43 |
| 142 | 44.02 | 0.01 | 0.69 | 0.09 | 13.86 | 1.05 | 0.33 | 7.70 | 22.95 | 0.47 | 91.24 |
| 143 | 23.17 | 0.03 | 3.36 | 0.32 | 45.20 | 0.19 | 2.40 | 16.77 | 0.09 | 0.49 | 92.25 |
| 144 | 26.60 | 0.07 | 2.47 | 0.27 | 41.28 | 0.21 | 1.53 | 18.29 | 0.13 | 0.63 | 91.63 |
| 145 | 29.96 | 0.06 | 1.21 | 0.29 | 38.51 | 0.23 | 0.52 | 18.56 | 1.01 | 0.47 | 90.92 |
| 146 | 28.63 | 0.03 | 2.11 | 0.29 | 40.53 | 0.25 | 1.50 | 17.54 | 0.88 | 0.46 | 92.36 |
| 147 | 23.49 | 0.08 | 2.84 | 0.32 | 43.89 | 0.20 | 3.40 | 15.37 | 0.10 | 0.25 | 90.20 |
| 148 | 30.36 | 0.07 | 1.47 | 0.52 | 34.15 | 0.20 | 0.38 | 21.65 | 0.29 | 0.52 | 89.72 |
| 149 | 38.39 | 0.06 | 1.96 | 0.28 | 24.99 | 0.27 | 0.64 | 14.60 | 10.12 | 0.55 | 91.94 |
| 150 | 29.92 | 0.07 | 1.05 | 0.27 | 40.47 | 0.24 | 0.61 | 19.48 | 0.06 | 0.38 | 92.68 |

Palme, Hezel, Ebel (2015, *EPSL*)

**Table S2:** Chemical composition of chondrules in Renazzo from Klerner (2001). FeO may be underestimated because metal was avoided in the analysis, see text for details.

|      | $SiO_2$ | $TiO_2$ | $Al_2O_3$ | $Cr_2O_3$ | FeO | MnO | MgO | CaO | total | Al/Ti |
|------|-------|-------|---------|---------|-------|------|-------|------|--------|------|
| Ch1  | 41.60 | 0.23  | 6.09    | 0.58    | 9.16  | 0.22 | 35.54 | 4.30 | 97.71  | 23.5 |
| Ch2  | 50.60 | 0.04  | 0.95    | 0.97    | 6.86  | 0.40 | 37.50 | 1.41 | 98.72  | 20.8 |
| Ch3  | 48.24 | 0.09  | 1.10    | 0.70    | 3.48  | 0.09 | 43.55 | 0.66 | 97.91  | 10.5 |
| Ch4  | 43.68 | 0.19  | 3.88    | 0.50    | 5.59  | 0.12 | 41.06 | 2.43 | 97.44  | 18.2 |
| Ch5  | 41.89 | 0.09  | 0.96    | 0.29    | 2.61  | 0.05 | 50.71 | 0.97 | 97.58  | 9.1  |
| Ch6  | 36.52 | 0.18  | 2.41    | 0.31    | 26.64 | 0.39 | 20.13 | 6.80 | 93.38  | 11.6 |
| Ch7  | 42.68 | 0.07  | 2.24    | 0.58    | 11.50 | 0.20 | 37.69 | 1.25 | 96.20  | 28.5 |
| Ch8  | 53.38 | 0.18  | 1.51    | 0.73    | 6.66  | 0.18 | 34.19 | 1.15 | 97.99  | 7.3  |
| Ch9  | 46.76 | 0.26  | 7.51    | 0.95    | 8.91  | 0.31 | 25.60 | 5.11 | 95.41  | 25.6 |
| Ch10 | 54.43 | 0.12  | 1.33    | 0.63    | 5.46  | 0.10 | 35.94 | 0.70 | 98.72  | 9.6  |
| Ch11 | 43.52 | 0.12  | 2.16    | 0.54    | 8.36  | 0.16 | 42.07 | 1.18 | 98.12  | 15.5 |
| Ch12 | 44.96 | 0.15  | 2.53    | 0.65    | 15.34 | 0.15 | 30.37 | 1.40 | 95.55  | 14.7 |
| Ch13 | 50.53 | 0.17  | 3.77    | 0.59    | 5.78  | 0.17 | 35.71 | 2.70 | 99.42  | 19.6 |
| Ch14 | 41.48 | 0.13  | 1.98    | 0.66    | 13.83 | 0.20 | 34.77 | 1.71 | 94.78  | 13.0 |
| Ch15 | 50.88 | 0.08  | 1.00    | 0.71    | 6.98  | 0.18 | 38.81 | 0.56 | 99.20  | 10.8 |
| Ch16 | 45.36 | 0.22  | 4.55    | 0.58    | 6.69  | 0.16 | 38.39 | 2.50 | 98.46  | 18.0 |
| Ch17 | 48.64 | 0.07  | 1.43    | 0.70    | 4.36  | 0.27 | 40.13 | 0.93 | 96.53  | 17.6 |
| Ch18 | 50.09 | 0.11  | 2.44    | 0.62    | 4.43  | 0.16 | 40.54 | 1.49 | 99.88  | 19.0 |
| Ch19 | 53.06 | 0.12  | 1.28    | 0.75    | 5.32  | 0.20 | 36.05 | 1.31 | 98.07  | 9.4  |
| Ch20 | 46.61 | 0.14  | 2.96    | 1.55    | 9.00  | 0.99 | 32.25 | 2.35 | 95.85  | 19.2 |
| Ch21 | 42.59 | 0.15  | 2.60    | 0.71    | 13.98 | 0.22 | 32.74 | 0.71 | 93.70  | 15.0 |
| Ch22 | 47.15 | 0.13  | 1.65    | 0.76    | 10.98 | 0.31 | 35.31 | 1.49 | 97.78  | 11.2 |
| Ch23 | 44.93 | 0.24  | 4.26    | 0.44    | 13.83 | 0.19 | 29.28 | 1.97 | 95.15  | 15.4 |
| Ch24 | 45.40 | 0.12  | 1.63    | 0.87    | 10.98 | 0.31 | 33.14 | 2.16 | 94.61  | 12.0 |
| Ch25 | 38.61 | 0.08  | 1.55    | 0.45    | 9.22  | 0.13 | 41.97 | 0.73 | 92.74  | 16.4 |
| Ch26 | 47.26 | 0.14  | 1.83    | 0.93    | 6.75  | 0.20 | 37.97 | 0.71 | 95.79  | 11.4 |
| Ch27 | 42.34 | 0.28  | 5.83    | 0.67    | 12.75 | 0.39 | 30.48 | 2.32 | 95.06  | 18.4 |
| Ch28 | 50.17 | 0.08  | 1.18    | 0.82    | 6.48  | 0.33 | 35.95 | 1.21 | 96.21  | 12.4 |
| Ch29 | 49.94 | 0.09  | 1.41    | 0.84    | 9.31  | 0.62 | 33.89 | 1.19 | 97.28  | 14.2 |
| Ch30 | 47.42 | 0.10  | 1.13    | 0.63    | 6.99  | 0.12 | 40.34 | 0.51 | 97.25  | 10.2 |
| Ch31 | 44.16 | 0.17  | 3.94    | 0.45    | 2.94  | 0.10 | 44.12 | 2.92 | 98.79  | 20.6 |
| Ch32 | 52.62 | 0.09  | 1.45    | 0.86    | 6.74  | 0.52 | 32.15 | 1.35 | 95.77  | 13.8 |
| Ch33 | 51.89 | 0.11  | 1.66    | 0.94    | 7.95  | 0.64 | 32.80 | 1.55 | 97.54  | 13.0 |
| Ch34 | 50.59 | 0.11  | 1.17    | 0.64    | 6.61  | 0.20 | 38.91 | 0.71 | 98.94  | 9.6  |
| Ch35 | 49.78 | 0.15  | 1.91    | 0.83    | 6.54  | 0.27 | 37.03 | 1.36 | 97.86  | 11.2 |
| Ch36 | 48.26 | 0.11  | 1.01    | 0.71    | 4.13  | 0.25 | 41.76 | 1.52 | 97.75  | 8.1  |
| Ch37 | 51.83 | 0.60  | 3.22    | 0.95    | 6.99  | 0.48 | 26.65 | 7.70 | 98.41  | 4.7  |
| Ch38 | 44.61 | 0.11  | 1.40    | 0.66    | 8.55  | 0.15 | 39.70 | 0.27 | 95.45  | 11.4 |
| Ch39 | 48.72 | 0.13  | 3.15    | 0.73    | 4.49  | 0.18 | 40.19 | 2.37 | 99.95  | 21.2 |
| Ch40 | 57.94 | 0.06  | 0.84    | 0.92    | 3.11  | 0.65 | 35.20 | 1.24 | 99.96  | 11.5 |
| Ch41 | 48.11 | 0.28  | 1.61    | 0.53    | 2.03  | 0.17 | 41.17 | 4.92 | 98.81  | 5.1  |
| Ch42 | 49.70 | 0.13  | 0.88    | 0.65    | 4.99  | 0.18 | 42.07 | 1.44 | 100.05 | 6.0  |
| Ch43 | 44.08 | 0.25 | 3.20 | 0.61 | 3.27 | 0.15 | 44.52 | 3.03 | 99.12 | 11.2 |
| Ch44 | 50.00 | 0.14 | 0.99 | 0.60 | 3.76 | 0.13 | 41.33 | 1.37 | 98.33 | 6.4 |
| Ch45 | 47.63 | 0.08 | 1.32 | 0.69 | 6.78 | 0.16 | 40.48 | 0.45 | 97.57 | 15.1 |
| Ch46 | 53.11 | 0.19 | 1.64 | 0.78 | 7.12 | 0.29 | 33.27 | 1.96 | 98.36 | 7.7 |
| Ch47 | 37.34 | 0.06 | 0.96 | 0.35 | 28.17 | 0.46 | 29.97 | 1.49 | 98.80 | 14.1 |
| Ch48 | 49.76 | 0.11 | 1.68 | 0.64 | 5.63 | 0.15 | 39.36 | 0.89 | 98.21 | 12.9 |
| Ch49 | 51.91 | 0.12 | 0.97 | 0.56 | 2.05 | 0.08 | 42.39 | 0.47 | 98.55 | 7.1 |
| Ch50 | 45.87 | 0.19 | 3.67 | 0.83 | 7.14 | 0.31 | 36.89 | 2.39 | 97.30 | 16.8 |
| Ch51 | 54.67 | 0.16 | 2.76 | 0.98 | 5.30 | 0.91 | 30.10 | 3.04 | 97.92 | 15.5 |
| Ch52 | 54.21 | 0.15 | 1.85 | 1.04 | 5.44 | 1.07 | 31.27 | 3.09 | 98.13 | 11.2 |
| Ch53 | 42.89 | 0.13 | 2.52 | 0.63 | 3.78 | 0.20 | 46.07 | 2.19 | 98.41 | 17.2 |
| Ch54 | 47.23 | 0.13 | 2.71 | 0.53 | 2.27 | 0.13 | 43.11 | 1.77 | 97.88 | 19.1 |
| Ch55 | 44.55 | 0.16 | 3.78 | 0.57 | 4.10 | 0.19 | 42.91 | 2.51 | 98.76 | 20.7 |
| Ch56 | 50.46 | 0.27 | 11.44 | 0.52 | 3.59 | 0.14 | 23.95 | 8.49 | 98.87 | 37.0 |
| Ch57 | 56.54 | 0.07 | 0.64 | 0.65 | 3.22 | 0.13 | 36.51 | 0.30 | 98.08 | 8.5 |
| Ch58 | 53.36 | 0.03 | 0.51 | 0.66 | 3.82 | 0.20 | 39.63 | 0.33 | 98.56 | 14.0 |
| Ch59 | 44.16 | 0.29 | 2.06 | 0.64 | 9.16 | 0.20 | 38.05 | 3.11 | 97.66 | 6.3 |
| Ch60 | 48.38 | 0.11 | 1.56 | 0.58 | 2.99 | 0.13 | 44.73 | 1.20 | 99.68 | 12.3 |
| Ch61 | 55.14 | 0.19 | 2.02 | 1.48 | 4.32 | 0.93 | 31.52 | 3.96 | 99.56 | 9.5 |
| Ch62 | 50.67 | 0.12 | 1.88 | 0.45 | 5.91 | 0.12 | 39.07 | 0.85 | 99.07 | 14.1 |
| Ch63 | 49.28 | 0.23 | 2.61 | 0.80 | 6.31 | 0.31 | 35.71 | 2.69 | 97.95 | 10.1 |
| Ch64 | 51.01 | 0.07 | 0.98 | 0.72 | 5.52 | 0.20 | 39.35 | 0.63 | 98.22 | 11.8 |
| Ch65 | 34.41 | 0.12 | 3.02 | 1.83 | 26.93 | 0.40 | 27.10 | 0.88 | 94.69 | 22.5 |
| Ch66 | 43.99 | 0.09 | 0.84 | 0.54 | 1.61 | 0.12 | 49.87 | 0.77 | 97.83 | 8.7 |
| Ch67 | 43.26 | 0.27 | 4.72 | 0.67 | 5.36 | 0.20 | 38.51 | 4.50 | 97.48 | 15.3 |
| Ch68 | 45.08 | 0.06 | 0.91 | 0.72 | 8.40 | 0.24 | 42.94 | 0.36 | 98.71 | 13.2 |
| Ch69 | 45.61 | 0.10 | 1.23 | 0.69 | 4.91 | 0.20 | 44.40 | 0.68 | 97.82 | 11.0 |
| Ch70 | 41.15 | 0.11 | 1.93 | 0.58 | 4.42 | 0.25 | 49.24 | 0.96 | 98.62 | 15.8 |
| Ch71 | 53.26 | 0.15 | 1.84 | 0.96 | 6.29 | 0.25 | 34.37 | 0.65 | 97.77 | 10.6 |
| Ch72 | 53.11 | 0.08 | 1.36 | 0.66 | 7.52 | 0.15 | 34.16 | 0.77 | 97.80 | 15.8 |
| Ch73 | 46.42 | 0.05 | 0.87 | 0.76 | 8.84 | 0.73 | 39.03 | 1.07 | 97.77 | 14.8 |